# Strong light-field effects driven by nearly single-cycle 7-fs light field in correlated organic conductors

[1]Yohei Kawakami, [1]Hirotake Itoh, [2]Kenji Yonemitsu, and [1]Shinichiro Iwai*

[1]Department of Physics, Tohoku University, Sendai 980-8578, Japan

[2]Department of Physics, Chuo University, Tokyo 112-8551, Japan

**Abstract** We have demonstrated transient charge localization effects with a driving high-frequency field of 7-fs, 1.5-cycle near infrared light in correlated organic conductors. In a layered organic conductor α-(BEDT-TTF)$_2$I$_3$ (BEDT-TTF: bis[ethylenedithio]-tetrathiafulvalene), a transient short-range charge order (CO) state is induced in a metallic phase. In contrast to such drastic change in the electronic state from the metal to the transient CO in α-(BEDT-TTF)$_2$I$_3$, dynamics of a field-induced reduction of a transfer integral are captured as a red shift of the plasma-like reflectivity edge in a quasi-one-dimensional organic conductor (TMTTF)$_2$AsF$_6$ (TMTTF: tetramethyltetrathiafulvalene). These studies on the field-induced charge localization have been motivated by the theory of dynamical localization on the basis of tight-binding models with no electron correlation under a strong continuous field. However, the results of pump-probe transient reflectivity measurements using nearly single-cycle 7-fs, 11 MV/cm pulses and the theoretical studies which are presented in this review indicate that the pulsed field contributes to the similar phenomenon with the help of a characteristic lattice structure and Coulomb repulsion.

## 1: Dynamical stabilization of electronic phases.

Stabilization in a periodically driven system has attracted much attention. One of examples is an inverted pendulum historically known as the Kapitza pendulum[1, 2] [Fig. 1(a)]. From a quantum mechanical point of view, a reduction of an ionization rate of an atom in the intense light-field has been discussed in terms of the analogous concept "adiabatic stabilization"[3, 4] which is analogous to the above classical system in the sense that the stabilization is induced by a strong oscillating field. On the other hand, realization of such a highly-nonequilibrium charge state with a strongly driving field has been difficult in solid state, because of heating effects which are caused by electron-electron scatterings and/or electron-phonon scatterings [5-12]. However, recent development of a few-cycle or a nearly single cycle < 10-fs , >10 MV/cm pulse enables us to induce a highly nonequilibrium charge state in solids. Such a transient charge state with a strongly driving field can open a new pathway toward light induced electronic phase transitions.

"Dynamical localization", i.e., the reduction of an effective transfer integral ($t_{eff} = \frac{1}{T_{field}} \int_0^{T_{field}} t e^{i\frac{e}{\hbar c} \mathbf{r}_{ij} \cdot \mathbf{A}(t')} dt'$) which is the time average of an original transfer integral $t$ (or equivalently an intersite hopping rate of charge) multiplied by the Peierls-phase factor ($\mathbf{A}(t) = -c\int_0^t \mathbf{E}(t')dt'$: vector potential for the applied electric field $\mathbf{E}(t)$, $r_{ij}$: position vector for corresponding transfer process, $T_{field}$: period of the oscillating field)[14-16] is a typical example of a periodically driven stabilization [Figs. 1(b) and 1(c)]. The dynamical localization has been originally developed for a continuous field, but similar effects are expected for a single-cycle pulse at least qualitatively. Considering that a transfer integral is one of important parameters which governs a charge motion in solids, related studies on strongly correlated systems are essential for aiming at a field-induced quantum phase transition[17-27].

Formation of a charge order (CO) is a well-known electronic phase transition, which is caused by the competition between the kinetic energy of charge (which is proportional to $t$) and the Coulomb repulsion energies (onsite $U$, intersite $V$). The CO has been a long lasting issue with respect to a metal-to-insulator (M-I) transition[28-30], superconductivity (SC)[31-33] and electronic ferroelectricity[34-36], since the discovery of the Verwey transition in magnetite[37]. Ultrafast melting of CO (or equivalently the ultrafast I-M transition) has been discussed in terms of a light-induced filling control and/or a structural change in strongly correlated materials such as transition metal oxides and low-dimensional organic salts [38-40].

In the progress of photoinduced dynamics in solid materials [38-53], electronic and/or structural orders such as CO[43-46], charge/spin density wave (CDW/SDW)[47-50], and SC[51-53] are optically modulated or even controlled and almost proceed to ultrafast bidirectional switching of phases. In contrast to well-understood "melting" of orders and/or "heating" effects [38-40], these modulations are achieved by enhanced Fermi surface nesting driven by ionic motion[48, 50], dynamic electron-phonon interaction[48] (CDW/SDW), phononic excitation[51, 52] and coherent excitation of a Higgs mode [53] (SC).

Here, we would like to describe transient formation of a short-range CO induced by a nearly single-cycle >10 MV/cm light-field[13]. Such a counterintuitive optical response is completely different from conventional melting of CO that is triggered by a photo-carrier generation [38-40] and is also different from the above mechanisms for CDW/SDW [48-50] and SC[51-53] modulations. In this review, such a characteristic mechanism is discussed on the basis of a polarization analysis of the transient CO[54], a reduction of the plasma frequency [55] in correlated organic conductors α-(BEDT-TTF)$_2$I$_3$ (BEDT-TTF: bis[ethylenedithio]-tetrathiafulvalene) and (TMTTF)$_2$AsF$_6$ (TMTTF: tetramethyltetrathiafulvalene), respectively. We choose these compounds as prototypical examples. The metal-to-insulator transition is accompanied by a drastic change in the optical conductivity spectrum, i.e., a spectral weight transfer from lower to higher energy regions in α-(BEDT-TTF)$_2$I$_3$. The near-infrared reflectivity spectrum that shows a plasma-edge-like shape is suitable for detecting the field-induced increase in *m* (or decrease in *t*) for (TMTTF)$_2$AsF$_6$. Both results are basically understood from reduced transfer integrals. However, their electronic dimensionality makes the outcomes quite different. α-(BEDT-TTF)$_2$I$_3$ is a quasi-two-dimensional material. The amplitude of the CO is large, so that their photoinduced change is also large in the reflectivity spectrum. (TMTTF)$_2$AsF$_6$ is a quasi-one-dimensional material. Quantum fluctuations substantially weaken the CO and justify the Lorentz analysis.

## 2: Experiment.

Reflection detected pump-probe (or equivalently transient reflectivity) measurements were performed by using both 7-fs and 100-fs light pulses. In 7-fs measurements, the 1.5-cycle pulses with a broad spectrum covering 0.56-0.95 eV were used as both pump and probe lights. The 7-fs probe pulse reflected from a sample was detected by an InGaAs detector after passing through a spectrometer. For 100-fs measurements, the central photon energy of the pump pulse and the probe range are 0.89 eV and 0.1-1 eV, respectively. The central photon energy of the pump light for the 7-fs measurement (~0.74 eV) is close to that of the 100-fs measurement (0.89 eV).

A setup for generating a < 10-fs pulse is schematically shown in Fig. 2(a). The broadband spectrum for the 7-fs pulse covers 1.3–2.2 μm (solid red line in Fig. 2(c)). This spectrum was obtained by focusing a carrier-envelope-phase (CEP) stabilized idler pulse (1.7 μm) from an optical parametric amplifier (Quantronix HE-TOPAS pumped by Spectra-Physics Spitfire-Ace) onto a hollow fiber (1 m) set within a Kr-filled (0.35 MPa) chamber (Femtolasers)[13, 54-60]. As shown by the circles, the spectral shape in Fig. 2 (c) is roughly reproduced by the calculated instantaneous angular frequency of the light pulse $\omega(t) = \omega_0 - \frac{n_2 \omega_0 z}{2c} |E_0|^2 \frac{d}{dt} e^{-\Gamma t^2}$ considering the nonlinear refractive index $n(t) = n_0 + \frac{1}{2} n_2 |E_0|^2 \exp(-\Gamma t^2)$ for the Gaussian light-field $E(t) = E_0 e^{-\Gamma t^2/2} e^{-i(\omega_0 t - kz)}$, where $n_0$, $n_2 \equiv 3\chi^{(3)}/(n_0^2 \varepsilon_0 c)$ =2.5x10$^{-19}$ m²/W for 0.35 MPa Kr, $E_0$, $\Gamma$, $\omega_0$, $k$, $z$ are the linear and third-order nonlinear refractive indices, amplitude of light-field, Gaussian parameter ($\Gamma = \frac{4\ln 2}{\Delta^2}$, $\Delta$ is the FWHM of the Gaussian), angular frequency, wavenumber $k = \frac{\omega_0 n(t)}{c}$, and coordinate along the propagating axis of light, respectively.

The time profile of this broadband light is characterized by an autocorrelation with a FWHM of 100 fs [solid line in Fig. 2(d)], which is much broader than the Fourier transform limit (~6 fs). This temporal broadening is caused by the propagation of light in an outlet quartz window (0.5 mm) of the chamber, a $CaF_2$ lens (1 mm) for collimation and dry air (5 m). The calculated autocorrelation profile is shown as circles in Fig. 2(d). Pulse compression was performed using both active mirror[60] and chirped mirror (Femtolasers) techniques. The optical setup for an active mirror compressor is shown in Fig. 2(b). The pulse width derived from the autocorrelation was 6 fs [Fig. 2(e)], which corresponds to 1.3 optical cycles, at the sample position, although we actually use 7-fs pulses to secure the stability and the intensity for pump-probe experiments.

In Sect. 4, the polarizations of both pump and probe pulses are controlled by a respective pair of wire-grid $CaF_2$ polarizers as shown in Fig. 3 [54]. The light intensity after the wire-grid polarizers is shown as a function of angles $\theta_2$ and $\theta_1$ in the figure. The pulse width derived from the autocorrelation was 7 fs at the sample position after the wire-grid polarizers and the window of the cryostat.

In the transient reflectivity measurements using the 7-fs pulses, the instantaneous electric field on the sample surface (excitation diameter of 200 μm) is evaluated as $E_{peak} = 9.8 \times 10^6$ (V/cm) for an excitation intensity $I_{ex}$ of 0.9

mJ/cm$^2$. Polarized optical reflection experiments were conducted on single crystals of α-(BEDT-TTF)$_2$I$_3$ (of size 0.5 × 0.7 × 0.2 mm) [61] and (TMTTF)$_2$AsF$_6$ (of size 1.5 × 0.7 × 0.2 mm) grown by electro-crystallization [62].

### 3: Transient short-range CO in α-(BEDT-TTF)$_2$I$_3$ [13].
#### a) Metal to insulator transitions in α-(BEDT-TTF)$_2$I$_3$

Figures 4(a) and 4(b) show the crystal structure of a layered organic conductor α-(BEDT-TTF)$_2$I$_3$ and charge distributions in CO insulator[63-65] and metallic phases. In this triangular lattice structure of α-(BEDT-TTF)$_2$I$_3$, the CO phase has crystallographically nonequivalent molecular sites A, A', B, and C [66-69]. Because of low symmetry (triclinic $P\bar{1}$), a charge imbalance between sites B and C remains even in the metallic phase [65, 66]. This charge imbalance is caused by nonequivalent transfer integrals in the low-symmetry structure, so that this should not be confused with any CO responsible for insulating states at low temperatures. We will discuss more in detail in Sect. 4. As illustrated in Fig. 4(b), this compound exhibits a thermal metal-to-CO insulator transition at $T_{CO}$ = 135 K [61, 64, 65] and a photoinduced transition between them. Efficient photoinduced I-M transitions of > 200 ET molecules/photon and an initial dynamics on an ultrafast time scale of ~20 fs have been reported[11, 70-80]. On the other hand, ultrafast photoinduced effects in the metallic phase above $T_{CO}$ had not been investigated because a strong light-field and fast time resolution are required to avoid heating. It is in contrast to the fact that the steady state has extensively been studied [81-86].

In the optical conductivity ($\sigma$) spectra in the mid- and near-infrared region [Fig. 5(a)], a clear opening of the CO gap at ~0.1–0.2 eV and a spectral weight transfer to a higher energy at the thermal metal-to-CO insulator transition[84-86] are detected. As shown in the steady state reflectivity ($R$) spectra at 140 K (metal: solid red curve) and 40 K (CO: dashed blue curve) [Fig. 5(b)], $R$ abruptly decreases at 0.09 eV [open circles in Fig. 5(d)] and increases at 0.64 eV [closed circles in Fig. 5(d)] as the temperature decreases below $T_{CO}$. Such a large change in $R$ (80% at 0.64 eV, 60% at 0.09 eV) at the metal-to-CO insulator transition directly corresponds to the opening of the CO gap and the transfer of the spectral weight to a higher energy of $\sigma$ [84-86]. The solid curve in Fig. 5(c) shows the spectral difference at different temperatures [$R$(40 K) - $R$(140 K)]/$R$(140 K), reflecting the metal-to-CO insulator change across $T_{CO}$. On the other hand, [$R$(190K) - $R$(140 K)]/$R$(140 K) (dashed line) and [$R$(170K) - $R$(140 K)]/$R$(140 K) (dashed-dotted line) exhibit increases in the electron-lattice temperature up to 190 K and 170 K. Thus, the marked increase in $R$ above 0.55 eV (blue shading) clearly characterizes the metal-to-insulator change across $T_{CO}$, while the rise in the electron-lattice temperature is detected as a reflectivity decrease below 0.65 eV.

b) Charge localization induced by 7-fs -light excitation

The transient reflectivity ($\Delta R/R$) spectra for time delays ($t_d$) of 30 fs (closed blue circles for an excitation intensity $I_{ex}$ = 0.8 mJ/cm², open blue circles for $I_{ex}$ = 0.12 mJ/cm$^2$) and 300 fs (closed black circles for $I_{ex}$ = 0.8 mJ/cm$^2$) after the excitation by a 7-fs pulse are shown in Fig. 6(a) [13]. The spectrum of the 7-fs broadband excitation pulse covering 0.57-0.95 eV is set on the higher-energy side of the main electronic band peaked at 0.1-0.2 eV for α-(BEDT-TTF)$_2$I$_3$ to avoid strong resonance. On this condition, it is reasonable to consider that non-resonant effects are dominant. The spectrum drawn with crosses represents $\Delta R/R$ at $t_d$ = 300 fs after a 100-fs pulse excitation for $I_{ex}$ = 0.8 mJ/cm$^2$. The blue shading in Fig. 6(a) exhibits a large increase in $R$ ($\Delta R/R \sim 0.28$ at 0.66 eV) at $t_d$ = 30 fs for $I_{ex}$ = 0.8 mJ/cm$^2$. The spectral shape of $\Delta R/R$ is analogous to that of the temperature differential spectrum [$R$(40K) - $R$(140 K)]/$R$(140 K), indicating that a photoinduced change from the metallic state to an insulating state. On the other hand, as shown by crosses, black closed circles, and red shading, the spectral shapes of $\Delta R/R$ measured at $t_d$ = 300 fs are analogous to [$R$(190K) - $R$(140K)]/$R$(140K), indicating a rise in the electron-lattice temperature up to 190 K. In addition, the $\Delta R/R$ spectrum for $I_{ex}$ = 0.12 mJ/cm$^2$, drawn with open blue circles, demonstrates that the metal-to-insulator change does not occur under a weak excitation.

c) **Charge ordering gap oscillation along time axis.**

Figure 6(b) shows a two-dimensional (probe energy– delay time) map of the $\Delta R/R$ spectrum [13]. The blue and red shadings represent increase and decrease in $R$ (positive and negative $\Delta R/R$), respectively. The spectrum for $t_d$ < 50 fs reflects the transient CO state. In this time domain, $\Delta R/R$ is modulated oscillatory with a period of 20 fs as shown by the red dotted lines. Then, the spectral shape of $\Delta R/R$ markedly changes reflecting the increase in the electron/lattice temperature represented by the red shading area for $t_d \sim 50$–100 fs. The spectrum is changed simultaneously as the oscillation decays.

The time evolution of $\Delta R/R$ observed at 0.64 eV is shown in Fig. 7(a)[53]. A positive $\Delta R/R$ (solid curve with blue shading) at $I_{ex} = 0.8$ mJ/cm$^2$ is observed for $t_d < 50$ fs. Then, $\Delta R/R$ becomes negative (red shading), indicating that the photoinduced CO state is melted because of the increase in the electron-lattice temperature. On the other hand, for $I_{ex} = 0.12$ mJ/cm$^2$, a positive signal was not detected at 0.64 eV, as shown by the dashed-dotted curve in Fig. 7(a). It is noteworthy that the time evolution of $\Delta R/R$ is modulated by the oscillating component with a period of 20 fs. The time-resolved spectrum of the oscillation at $t_d = 0$–40 fs analyzed by the wavelet (WL) transformation [blue curve, inset of Fig. 7(b)] corresponds to the $\sigma$ spectrum near the CO gap of ~0.1 eV at 10 K(black curve in the inset). Therefore, this oscillation is attributed to the intermolecular charge oscillation reflecting the CO gap.

Thus the large reflectivity change of > 25% and the coherent charge oscillation along the time axis reflect the transient opening of the short-range CO gap in the metallic phase.

### d) Mechanism of charge localization

In states driven by a strong AC field $E(\omega)$ far from an equilibrium one, a transfer integral $t$ is effectively reduced for continuous[14-18] and pulsed[17-21] light-fields. According to the dynamical localization theory[14-16] for a continuous AC field, the effective $t$ ($= t_{eff}$) governing long-time behaviors is given by

$$t_{eff} = t_0 J_0 (\Omega_{AB} / \omega), \qquad (1)$$

where $t_0$ is the transfer integral in equilibrium and $J_0$ is the zeroth-order Bessel function. Here, $\Omega_{AB} \equiv e\bm{r}_{A,B} \cdot \bm{E}_0 / \hbar$, with $\bm{E}_0$ and $\omega$ being the amplitude and angular frequency, respectively, of the electric field of the light; $\bm{E}(t) = \bm{E}_0 \sin(\omega t)$; $\bm{r}_{A,B}$ is the vector from site A to its nearest-neighbour site B, which are the centers of the corresponding molecules shown in Fig. 4(b); and $e$ is the elementary charge. Although Eq. (1) is basically satisfied for continuous light fields in the original theory[14-16], this relation also holds for pulsed light fields[20] with respect to the efficiency of the intersite electronic transition (i.e., the energy increment due to the pulsed light). In addition, DMFT calculations show that the change in the electronic state into one with modified $t$ appears within the time scale of a few optical cycles after the sudden application of a continuous AC field[17, 18]. These facts suggest that the dynamical localization transiently functions with few-cycle pulses. If we use the parameter $r_{AB}\cos\theta$ =5.4 angstroms[66]; where $\theta$ represents the relative angle between $\bm{r}_{A,B}$ and $\bm{E}$, the effective transfer integral $t'$ becomes zero at 40 MV/cm ($\Omega_{AB}/\omega = 2.40$), as shown in Fig. 1(c). According to Fig. 1(c), the change in $J_0 \propto t'$ is estimated to be approximately 10% for a typical instantaneous field of 9.3 MV/cm ($I_{ex}$ = 0.8 mJ/cm²).

To consider a possible instability of the metallic phase induced by the 10% change in $t$, the change in $T_{CO}$ is roughly estimated. The hole densities ($\rho_H$) at sites A and A' [Fig. 4(b)] are plotted as a function of the normalized temperature $T/T_{CO}$ in Fig. 8, using the Hartree–Fock approximation for an extended Hubbard model[67]. The closed (A: charge rich) and open (A': charge poor) circles correspond to the original $t_0$, where a charge disproportionation occurs below $T/T_{CO}$ = 1. The hole densities $\rho_H$ as a function of $T/T_{CO}$ are calculated for $t$'s equal to $0.95 t_0$(rectangles), $0.9 t_0$(upward triangles), and $0.85 t_0$(downward triangles). This indicates that $T_{CO}$ increases across the measured temperature (138 K).

Thus, the reduction of effective transfer integrals is about ~10% for 10 MV/cm according to the dynamical localization theory. It is found that the contributions from $U$ and $V$ are very important. In the next section, more detailed analyses are performed based on a polarization analysis and exact diagonalization studies for photoinduced real-time dynamics [54].

## 4. Polarization selectivity of charge localization in an organic metal [54]
### a) Anisotropic triangular structure of α-(BEDT-TTF)$_2$I$_3$.

Figure 9(a) shows the triangular lattice structure of α-(BEDT-TTF)$_2$I$_3$ with crystallographically nonequivalent molecular sites A, A', B, and C which are linked by $b_1(b_1')$-$b_4(b_4')$, $a_1(a_1')$, $a_2$ and $a_3$ bonds in the CO phase [66-69]. A 1010-type CO is formed on sites A and A' along $a$ ($a_2$ and $a_3$) bonds, which constitutes charge-rich A-B-A-(black) and charge-poor A'-C-A'-(white) zig-zag stripes. As shown in Fig. 9(b), sites A and A' are equivalent above $T_{CO}$ (metallic phase) as bonds $b_1$-$b_4$ and $b_1'$-$b_4'$ are and bonds $a_1$ and $a_1'$ are. However, already in the *metallic* phase, site B is charge-rich because site B is located between two $b_2$ bonds that possess the largest transfer integral, while site C is charge-poor because site C is located between two $b_1$ bonds that possess the second largest transfer integral. Thus, this charge imbalance between sites B and C originates from the kinetic term of the Hamiltonian (i.e., the kinetic energy) [65, 66] and has nothing to do with the CO in the insulating phase at low temperatures. The intersite interactions that produce the 1010-type CO along $a$ bonds are responsible for the insulating ground state. In equilibrium, the CO only appearing in the insulating phase should not be confused with the charge imbalance that appears in both phases.

In this section (Sect.4), we describe the dependence of $\Delta R/R$ on polarizations of both 7-fs pump- and probe- pulses (see Sect. 2 and Fig. 3). Our result indicates that a short-range CO along the $a_2$ and $a_3$ bonds is induced efficiently for the pump with polarization perpendicular to the 1010-CO axis. This is in contrast to the intuitive expectation.

Steady state $R$ spectra measured for various polarizations at 138 K (metallic phase) are shown in Figs. 10(a) and 10(b). Figure 10(c) shows the spectral differences between 138 K (metal) and 50 K (insulator) $(\Delta R/R)_{MI} = \{R(50K) - R(138K)\}/R(138K)$. Here, the angle $\theta$ is between the $b$ axis and the light polarization, as shown in Fig. 3. The solid red line in Fig. 10(c) clearly shows that $R$ for > 0.6 eV markedly increases upon the metal-to-CO insulator transition for $\theta = 0°$ (//$b$), whereas $(\Delta R/R)_{MI}$ is almost zero for $\theta = 90°$ (//$a$) (black two-dotted line). Thus, the polarization dependence of $(\Delta R/R)_{MI}$ at >0.6 eV is recognized as the fingerprint of the CO caused by charge motion mainly along the $b$ bonds.

### b) Dependence of $\Delta R/R$ on polarization of probe pulse

In Fig. 10(c), $\Delta R/R$ spectra for $\theta_{pr}$ (angle between the polarization of the probe pulse and the $b$ axis as shown in Fig. 3)=0° (//$b$) and =90° (//$a$) for $t_d$ = 30 fs at 138 K are shown by the red-filled circles and the blue squares, respectively. Here, $\theta_{pu}$ (angle between the polarization of the pump pulse and the $b$ axis as shown in Fig. 3)=0° (//$b$) for both spectra. The $\Delta R/R$ spectrum for $\theta_{pr}$=0° (//$b$) shows a large (>25 %) increase in $R$ at 0.67 eV, while the $\Delta R/R$ for $\theta_{pr}$=90° (//$a$) is small (~1%). Such polarization dependence, which takes a maximum at $\theta_{pr}$=0° agrees well with the fingerprint of the CO $(\Delta R/R)_{IM}$ at 0.65 eV.

### c) Dependence of Δ*R*/*R* on polarization of pump pulse

The $\theta_{pu}$ dependence of $\Delta R/R$ ($t_d$=30 fs) at 138 K is shown in Fig. 11 for various $\theta_{pr}$ (0º, 23º, -23º, 90º). The short-range CO is observed to be efficiently induced for $\theta_{pu}$ =0º (//*b*) for all $\theta_{pr}$, whereas it is almost unaltered for $\theta_{pu}$ =90º (//*a*). In fact, $\Delta R/R$ for $\theta_{pu}$ =0º is >10 times larger than that for $\theta_{pu}$ =90º. Such a drastic change of $\Delta R/R$ depending on $\theta_{pu}$ is not attributed to the dependence of the penetration depth *d* on $\theta$ because the variation of *d* is smaller than twice for //*a* and //*b* (~$10^{-4}$ cm for //*a*, 5x$10^{-5}$ cm for //*b* in our detection range). Thus, the result in Fig. 11 is not affected so seriously by the polarization dependence of *d*. A plausible reason why the CO is efficiently induced for $\theta_{pu}$ =0º (//*b*) is that the reduction of $t_{eff}$ in the $b_1$ and $b_2$ bonds during the nearly single-cycle pulse is responsible for the CO.

### d) Theoretical consideration using time-dependent Schrödinger equation [23, 54].

To consider a mechanism from the microscopic viewpoint, a numerical calculation is performed for the strong light-field effect using a time-dependent Schrödinger equation for the 1/4-filled two-dimensional extended Hubbard model[23, 54],

$$H = \sum_{\langle i,j \rangle \sigma} t_{ij} \left( c^+_{i,\sigma} c_{j,\sigma} + c^+_{j,\sigma} c_{i,\sigma} \right) + U \sum_i n_{i\uparrow} n_{i\downarrow} + \sum_{\langle i,j \rangle} V_{ij} n_i n_j \quad , \qquad (2)$$

where $c_{i\sigma}$ is the annihilation operator of a hole on site *i* with spin σ,

$n_{i,\sigma} = c_{i\sigma}^{+} c_{i\sigma}$, and $n_i = \sum_{\sigma} n_{i,\sigma}$ [23]. This model has on-site Coulomb repulsion ($U$), nearest-neighbor repulsions ($V_{ij}$) and transfer integrals between sites $i$ and $j$ ($t_{ij}$). Here, we employ the parameters for the metallic phase. We used the exact diagonalization method for systems consisting of 16 sites with periodic boundary conditions. The transfer integrals $t_{ij}$ had been evaluated by the extended Hückel calculation using the X-ray structural analysis at the metallic phase[66]. The coupling with electric fields is introduced through the Peierls phase. The time-dependent Schrödinger equation is numerically solved during and after the excitation of a single-cycle pulse with central frequency $\hbar\omega$ = 0.8 eV. The change in correlation functions are calculated: the spatially and temporally averaged double occupancy $\langle\langle n_{i\uparrow} n_{i\downarrow} \rangle\rangle$ and the averaged nearest-neighbor density-density correlations $\langle\langle n_i n_j \rangle\rangle_{a2,a3}$ for the $a_2$ and $a_3$ bonds and $\langle\langle n_i n_j \rangle\rangle_b$ for the $b$ bonds. Nearest-neighbor correlations show how the short-range 1010-CO is enhanced along the $a_2$ and $a_3$ bonds. Here, temporal averages were calculated for $5 T_{\text{field}} < t_d < 10 T_{\text{field}}$ with $T_{\text{field}}$ being the period of the oscillating electric field. The present finite-size-system calculation cannot spontaneously break the symmetry or produce a long-range CO. However, it is useful for discussing the generation of a short-range CO.

The averaged correlation functions $\langle\langle n_{i\uparrow} n_{i\downarrow} \rangle\rangle$, $\langle\langle n_i n_j \rangle\rangle_{a2,a3}$ and $\langle\langle n_i n_j \rangle\rangle_b$ are shown in Figs. 12(a)-(c) as functions of $eaF/\hbar\omega$ ($a$: component parallel to the $b$ axis of the molecular spacing, $F$: field amplitude) for $U$=0.8 eV, intersite

Coulomb repulsions $V_1$(along all $b$ bonds)=0.3 eV, $V_2$ (along all $a$ bonds) =0.35 eV. The quantity $\langle\langle n_{i\uparrow}n_{i\downarrow}\rangle\rangle$ decreases with increasing $eaF/\hbar\omega$ for <0.4 [Fig. 12(a)], which means that holes with opposite spins avoid being on the same site more strongly as if $U$ were transiently increased relative to the transfer integrals. The decrease in $\langle\langle n_{i\uparrow}n_{i\downarrow}\rangle\rangle$ (14 % for $eaF/\hbar\omega$ =0.37) could be reproduced in the ground state, if we increased $U$ by 6.5%. The decrease in $\langle\langle n_i n_j\rangle\rangle_{a2,a3}$ (15 % for $eaF/\hbar\omega$=0.37) [Fig. 12(b)] means that holes avoid neighboring along the $a_2$ and $a_3$ bonds more strongly as if the $V_{ij}$ along the $a_2$ and $a_3$ bonds were transiently increased relative to $t_{ij}$.

The $\theta_{pu}$ dependence of the averaged nearest-neighbor density-density correlation for the $a_2$ and $a_3$ bonds $\langle\langle n_i n_j\rangle\rangle_{a2,a3}$ is shown in Fig.12(d) for $eaF/\hbar\omega$=0 (black crosses), 0.14 (blue rhombuses), 0.20 (green triangles), 0.27 (red squares) ($U$= 0.8 eV, $V_1$=0.3 eV, $V_2$=0.35 eV). These data reflect the 1010-type CO along the $a_2$ and $a_3$ bonds. The reduction of $\langle\langle n_i n_j\rangle\rangle_{a2,a3}$ is most efficient at $\theta_{pu}$ =0~10º for any $eaF/\hbar\omega$, which is almost parallel to the $b$ axis. This polarization dependence of $\langle\langle n_i n_j\rangle\rangle_{a2,a3}$ is consistent with that of $\Delta R/R$ [Fig. 11]: the efficiency of strengthening the short-ranged CO is maximized by polarization along the $b$ axis within the accuracy of $\theta_{pu}$.

The comparable Coulomb repulsions $V_2$ and $V_1$ in the characteristic low-symmetric molecular backbone (i.e., large and competing $V_2$, $V_1$, and $t_{b2}$ effects) are essential for the transient charge localization, as described in the original papers[23, 54]. This fact is consistent with the fact that any dramatic localization has not been detected in a quasi 1-D organic metal (TMTTF)$_2$AsF$_6$ [55] as described in the next section (Sect. 5). The correlation-supported mechanism of the transient short-range CO in the present study is conceptionally different from the mechanisms for light-induced CDW[48, 49] and SC[51, 52]. Thus, a new strategy for light-induced control of correlated electron materials may be opened.

In summary of this section (Sect. 4), the dependence of the transient reflectivity spectra on polarizations of 7-fs pump and probe lights were investigated for clarifying the controllability and the mechanism for transiently inducing the short-range CO in an organic metal α-(BEDT-TTF)$_2$I$_3$. Efficient induction of the short-range CO for the pump polarization perpendicular to the 1010-CO axis is realized by re-distribution of charges through the *b* bonds and competing intersite Coulomb interactions along the *a* and *b* bonds in the triangular lattice.

5. Ultrafast response of plasma-like reflectivity edge in (TMTTF)$_2$AsF$_6$ driven by strong-light field[55]

a) Plasma-like reflectivity edge in (TMTTF)$_2$AsF$_6$

As shown in the previous sections (Sect. 4 and 5), we have succeeded at realizing a transiently charge-localized state utilizing 7-fs nearly single-cycle pulses in α-(BEDT-TTF)$_2$I$_3$ [13, 54]. Such a light-field-induced short-range CO has been originally motivated by the dynamical localization, i.e., the increase in the effective mass *m* or the decrease in *t*. However, the direct detection of *m* on a short time scale had not yet been performed. Here, we demonstrate a reduction in *t* (or equivalently an increase in *m*) by measuring a spectral change in a "plasma-like" reflectivity edge.

(TMTTF)$_2$AsF$_6$ (Fig. 13) is a typical quarter-filled organic conductor [34, 62, 87-89], showing CO below $T_{CO}$=102 K. However, the correlation effect is generally believed to be rather weak in comparison with that for α-(BEDT-TTF)$_2$I$_3$. Such a small-gap (0.1–0.2 eV) insulator has a near-infrared (~0.7 eV) reflectivity edge similar to the plasma edge of metals [87, 89, 90] as shown in Fig. 14(b). This plasma-like reflectivity edge can be characterized by $\omega_p = \sqrt{ne^2/(\varepsilon_\infty \varepsilon_0 m)}$ in the Lorentz model, if $\omega_p \gg \omega_0$ (number of charges $n$: ~2 x 10$^{21}$cm$^{-3}$ in the 1/4 filled-band and their mass *m*: 3–4$m_0$, permittivities for high-frequency and vacuum $\varepsilon_\infty, \varepsilon_0$, charge gap for transverse wave $\hbar\omega_0$=0.1–0.2 eV)[91]. In this sense, we refer to this reflectivity edge as a "plasma-like edge", because it reflects the collective charge motion much above the gap. In this section (Sect. 5), we describe the field-induced change in *m* or *t* probed by the spectral change in the plasma-like reflectivity edge around $\hbar\omega_p$ ~0.7 eV($\gg \hbar\omega_0$).

### b) Steady state reflectivity of (TMTTF)$_2$AsF$_6$

The $\sigma$ and *R* spectra of (TMTTF)$_2$AsF$_6$ at 25 and 150 K for the polarization parallel to *a* (*E*//*a*) are shown in Figs. 14(a) and 14(b), respectively, where *a* is the stacking axis of the TMTTF molecules. The *R* spectra [Fig. 14(b)] have a plasma-like edge even below $T_{CO}$ because of the relation $\omega_p \gg \omega_0$. The existence of the CO gap is unclear in this spectral range at $T \lesssim T_{CO}$, while the $R$ spectra at > 0.2 eV can be well reproduced using the Lorentz model (solid lines) that takes account of vibrational couplings [87, 89, 91]. The discrepancy between the data and the fitted curve for 0.1-0.2 eV(corresponding to vibrational peaks) at 25 K is caused by the dielectric screening in the vibrational response by the charge [91]. The fitting parameters are $\hbar\omega_p$ =0.703 eV with the scattering frequency $\gamma$ =0.125 eV(25 K) and 0.154 eV(150 K) and $\hbar\omega_0$ = 0.180(25 K) and 0.193 eV(150 K). These parameters [91] are consistent with those of another TMTTF salt with a different anion $(TMTTF)_2PF_6$ at 300 K in the earlier study [87]. The spectral shape around the reflectivity edge (~0.7 eV) is governed by $\omega_p$ and $\gamma$, i.e., the gap structure in the low energy (< 0.2 eV) spectrum[87, 91] does not seriously affect the spectrum around the plasma-like edge.

The temperature dependence of $\gamma$ as $\Delta\gamma/\gamma_{25K} = (\gamma - \gamma_{25K})/\gamma_{25K}$ is shown in the inset of Fig. 14(b). Note that $\gamma$ does not depend on the temperature below $T_{CO}$, and it abruptly starts to increase at $T_{CO}$ with increasing temperature. Because the parameter $\gamma$ in the Lorentz model represents the scattering frequency, this anomalous temperature dependence of $\gamma$ should be caused by the temperature dependence of electron–electron scatterings, i.e., the frozen charges at $T \lesssim T_{CO}$ do not contribute to change $\gamma$, while the mobile charges at $T > T_{CO}$ should increase $\gamma$ with increasing temperature.

c) Red-shift of plasma-like reflectivity edge: Transient reflectivity measured by 100-fs pulse

The temporal change in the $\Delta R/R$ spectrum of $(TMTTF)_2AsF_6$ (15 K) for $t_d$=0-4 ps measured using a 100-fs pulse (excitation intensity $I_{ex}$ = 0.5 mJ/cm$^2$, 2 MV/cm)[55] is represented by the two-dimensional (probe-energy -delay time) map in Fig. 15(a). The polarizations of the pump and the probe pulses ($E_{pu}$ and $E_{pr}$, respectively) are parallel to the $a$-axis ($E_{pu}//a$, $E_{pr}//a$). The excitation energy 0.89 eV, where the penetration depth is very large (~1 μm), was chosen for nearly non-resonant strong field application.

The $\Delta R/R$ spectrum at $t_d$ = 0.1 ps shown in Fig. 15(b) is well reproduced by the Lorentz analysis (orange curve) on the assumption of a 1.8 % decrease in $\omega_p$, a 12 % increase in $\gamma$ and a 11 % increase in the width of the vibrational peak at 0.165 eV [55, 91] in the Lorentz model. The spectral change calculated with only decreasing $\omega_p$ (blue line) and that with only increasing $\gamma$ (red line) are shown in Fig. 15(c), which enables us to distinguish the contributions from $\Delta\omega_p/\omega_p$ (blue curve) and $\Delta\gamma/\gamma$ (red curve) easily. Note that a ~2 % reduction in $\omega_p$ can be detected as a very large (~30 %) change of $\Delta R/R$ in the spectral range of $\omega_p$. On the other hand, the 12 % increase in $\gamma$ indicates that the temperature increases up to ~120 K from 15 K across $T_{CO}$ at $t_d$=0.1 ps.

### d) Ultrafast reduction of $\omega_p$: Transient reflectivity measured by 7-fs pulse

The temporal change in $\Delta R/R$, for $t_d$ = 0–150 fs, measured using the 7-fs pulse ($I_{ex}$ = 0.8 mJ/cm², 9.8 MV/cm$^{-1}$) is shown in Fig. 16(a). The spectral window 0.58–0.95 eV in Fig. 16(a) is shown by the white arrow in Fig. 15(a). A quantitative comparison between the results of 100-fs and 7-fs measurements is difficult because the excitation pulse energy and the field amplitude of the 7-fs pulse are higher than those of the 100-fs pulse. However, the spectral shapes of $\Delta R/R$ for these measurements are qualitatively consistent at $t_d$ = 0.1 ps, although the detailed values of $\Delta\omega_p$ and $\gamma$ are different, as described below. The time evolutions of $\Delta R/R$ at i) 0.85, ii) 0.73 and iii) 0.62 eV are shown in Fig. 16(b). The decrease in $R$ appears on the time scale of ~20 fs, and has an oscillating structure with a period of 20 fs at 0.85 eV, reflecting the ultrafast reduction and the oscillation of $\omega_p$ [Fig. 16(b) (i)]. The slower rise (~80 fs) is observed at 0.62 eV [Fig. 16(b) –(iii)], where $\Delta R/R$ reflects the increase in $\gamma$ [Fig. 15(c)].

Figures 17(a)-17(f) show the $\Delta R/R$ spectra at various time delays, $t_d$ = 0–80 fs. As indicated by the blue curve in Fig. 17(b), $\Delta R/R$ ~-0.4 at around 0.7 eV corresponds to the 2.8% decrease in $\omega_p$ at $t_d$= 18 fs, whereas a 30% increase in $\gamma$ at $t_d$ =80 fs [red-dashed curve in Fig. 17(f)] with a 1.7% decrease in $\omega_p$ [blue-dashed curve in Fig. 17(f)] is observed. Thus, the ultrafast (~20 fs) reduction of $\omega_p$ and the slower (<80 fs) increase in $\gamma$ corresponding to the slow electron temperature increase are demonstrated.

In general, the decrease in $\omega_p$ corresponds to an increase in $m$ or a decrease in $n$. Here, $n$ is the number of the charges in the 1/4 filled-band system, i.e., charge 0.5 e per TMTTF molecule on average. Thus, in the present system, the number of these charges cannot decrease upon excitation with the photon energy well below any interband transition from the present 1/4 filled-band. In fact, the 2.8 % decrease in $\omega_p$ (5.8% increase in $m$) is consistent with a ~10% decrease in $t$ estimated in α-(BEDT-TTF)$_2$I$_3$ [13] suggested in the previous section (Sect.3).

### e) Dynamics of charge localization and electronic thermalization.

As shown by the green shade in Figs. 17(d) and 17(e), a spectral dip in the $\Delta R/R$ spectra at ~0.7 eV is observed at $t_d$ = 35 and 50 fs. They cannot be understood in the framework of the Lorentz model. Electron–electron or electron–phonon scatterings in such an early time region are not described by $\gamma$, i.e., a picture based on stochastic processes is not valid because i) scattering occurs only once or twice for $\hbar/\gamma$ of about 40 fs, and ii) the coherence of charges induced by the oscillating light-field can survive at such an early stage. Thus, we should employ a deterministic description using an interaction between the oscillating charges with frequency $\omega_p$ and other charges and/or lattice, instead of a stochastic description.

In the simplest model of elastic scattering, the phase of oscillating charges merely shifts. Such a deterministic picture is justified during the short time period while few scatterings occur. Then, a typical response function $\chi(t)$ for the oscillating charges is schematically shown in Fig. 18(a). The spectral response function $\chi(t)$ is obtained by the Fourier transform of $\chi(t)$. The windowed Fourier transforms lead to the spectrogram of $\chi(t)$ [Fig. 18(b)] and time resolved spectra $\chi(t)$ at 20 fs (black line) and 40 fs (red line) [Fig. 18(c)]. The phase shift results in a spectral dip [Fig. 18(b)]. Thus, one possible explanation for the spectral dip is an interaction between the charges oscillating with $\omega_p$ and other charges or lattice modes. To discuss the origin of the experimentally observed spectral dip, we should make further investigations from both experimental and theoretical viewpoints. However, they are beyond the scope of this review.

The time evolutions of $-\Delta\omega_p/\omega_p$ (blue dots) and $\Delta\gamma/\gamma$ (red dots) are obtained from the Lorentz analysis and shown in Fig. 19(a). Figure 19(b) shows the time evolution of the green shaded area of the dip in Figs. 17(d) and 17(e) discussed above. The dip appears at $t_d = $ ~40 fs, which approximately corresponds to the averaged scattering time $\hbar/\gamma$ for the steady state $R$ spectrum [$\gamma = 0.125$ eV(25 K)], although the mechanism of this agreement remains unclear. In addition, the decay time of the dip roughly corresponds to the growth time of $\gamma$, which seems to reflect the crossover from a coherent process to a stochastic process.

An oscillating structure with a period of 20 fs is observed in the time evolution of $-\Delta\omega_p/\omega_p$ in Fig. 19(a). This 20-fs oscillation is also seen in the raw data in Figs. 16(a) and 16(b)–i). The period corresponds to the $\sigma$ peak at

~0.2 eV in Fig. 14(a), corresponding to the charge gap ($\hbar\omega_0$=0.1~0.2 eV) [62, 92]. The different energy scales between $\hbar\omega_p$ (0.7 eV) and $\hbar\omega_0$ (0.1~0.2 eV) allow us to detect the time-domain oscillation of $\omega_p$ with frequency corresponding to $\omega_0$.

It is finally noted that the $-\Delta\omega_p/\omega_p$ persists even after ~ps. We observed a coherent phonon with a frequency of 62 cm$^{-1}$ for the longer time delay (not shown) which has been assigned as an alternating displacement for the direction perpendicular to the one-dimensional chain [89]. Considering that the 62 cm$^{-1}$ mode strongly modulates *t*, the long-lived $-\Delta\omega_p/\omega_p$ is attributable to the lattice stabilization. Owing to this lattice effect, the reduction of *t* persists even after the rise in the electron temperature. This is very different from the fact that the transient CO disappears within 40 fs in α-(BEDT-TTF)$_2$I$_3$ [13].

## 6. Summary & future perspective

We have demonstrated that transient charge localization is induced by a driving high-frequency field of a 7-fs near infrared 1.5-cycle pulse in organic conductors. In the quasi two-dimensional system α-(BEDT-TTF)$_2$I$_3$, the transient short-range CO state in the metallic phase is realized. In contrast to such drastic change from the metal to the CO in α-(BEDT-TTF)$_2$I$_3$, the dynamics of the field-induced reduction in the transfer integral is captured as the red shift of the plasma-like reflectivity edge in the quasi-one-dimensional organic conductor (TMTTF)$_2$AsF$_6$.

Our studies on the charge localization have been motivated by the theory of dynamical localization for tight-binding models under a continuous field [14-18]. However, the results of pump-probe experiments using 1.5- cycle 7-fs pulses and the theoretical studies which are presented in this review indicate that a pulsed field contributes to the localization with the help of their characteristic lattice structures and Coulomb repulsion.

Here, we summarize the difference between the results of α-(BEDT-TTF)$_2$I$_3$ [13, 54] (Sect. 3, 4) and those of (TMTTF)$_2$AsF$_6$ [55] (Sect. 5). In the quasi- two-dimensional organic conductor α-(BEDT-TTF)$_2$I$_3$, the reduction of $t$ just above $T_{CO}$ can induce the metal-to-insulator transition accompanied by the drastic change in the optical conductivity spectrum, i.e., spectral weight transfers from lower to higher energy regions. On the other hand, in (TMTTF)$_2$AsF$_6$, the charge disproportionation [0.585(rich)-0.415(poor)] in the CO phase is much smaller than that in α-(BEDT-TTF)$_2$I$_3$ [0.8(rich)-0.2(poor)]. This can be understood by the fact that quantum fluctuations are more effective to reduce the long-range order in quasi-one-dimensional (TMTTF)$_2$AsF$_6$. Therefore, optical excitations in the near-infrared region are little influenced by the CO in (TMTTF)$_2$AsF$_6$. Indeed, the near-infrared reflectivity at low temperature is well reproduced by the Lorentz model characterized by $\omega_p$ and $\gamma$, while the weak bond alternation is responsible for the dimerization gap [92] at 0.1~0.2 eV. Therefore, the near-infrared spectrum of (TMTTF)$_2$AsF$_6$ is suitable for detecting the field-induced increase in $m$ (or decrease in $t$), although this compound is not suitable for seeking any field-induced drastic transition [as actually observed in α-(BEDT-TTF)$_2$I$_3$].

As mentioned above, the theories of the dynamical localization have been originally developed for a continuous field with a frequency much higher than resonant frequencies. As for the usage of a pulsed field, this concept is (at least qualitatively) valid also for a 1.5-cycle pulse, because the essence of this effect is the time average during the cycle (~5 fs for near infrared region) of the oscillating field as shown in the introduction ($t_{eff} = \frac{1}{T_{field}} \int_0^{T_{field}} t e^{i\frac{e}{\hbar c}\mathbf{r}_{ij}\cdot\mathbf{A}(t')} dt'$).

In such a short-time domain, it is difficult to include contributions from resonance effects. In fact, the central photon energy of our excitation pulse is ~0.8 eV which is not so far from the resonance that could be realized for a very long pulse. Thus, resonance effects are not so effective for a nearly single-cycle field in comparison with those for a continuous field, although we should consider those more in detail for quantitative studies.

Another interesting point is that the transient CO persists for ~40 fs after the electric field of light turns off in α-(BEDT-TTF)$_2$I$_3$. In contrast to the longer lifetime (~1 ps) of the reduction of $t$ in (TMTTF)$_2$AsF$_6$ realized by the lattice stabilization, the 40-fs lifetime in α-(BEDT-TTF)$_2$I$_3$ suggests a different possibility, i.e., the nonequilibrium electronic state driven by the light-field survives for ~40 fs as suggested by theories considering the correlation effect [23, 24].

As a future problem, application of a half-cycle (~2 fs) light-field is expected to cause a drastic response. The situation is very different for the half-cycle (~2 fs) pulse from the ~single-cycle application shown in this review. If we optimize a CEP for the half-cycle pulse, we will be able to realize a direct manipulation of an electronic polarization in the ultrafast peta-Hz (PHz) frequency region. Considering that our target compounds α-(BEDT-TTF)$_2$I$_3$ and (TMTTF)$_2$AsF$_6$ are recognized as electronic ferroelectrics, a PHz manipulation of the electronic ferroelectricity is expected.

In this review, we focused on the strongly correlated metal or the small gap insulator. Another interesting target for the strong light-field-effect in correlated systems is superconductors. From a theoretical point of view, a new mechanism of photoinduced superconductivity is discussed in terms of the dynamical localization[26]. Furthermore, investigation of strong field effects on superconductivity in femtosecond - attosecond time scales enables us to expect clarification of mechanisms of high-temperature superconductivity with energy scales > 0.1 eV[93].

Acknowledgements

We thank T. Ishikawa, Y. Naitoh, Y. Sagae, Y. Yoneyama, T. Amano (Tohoku Univ.), M. Dressel (Universität Stuttgart), K. Yamamoto (Okayama Univ. of Sci.), T. Sasaki (IMR, Tohoku Univ.), Y. Nakamura, H. Kishida (Nagoya Univ.), S. Ishihara (Tohoku Univ.), Y. Tanaka(Chuo Univ.) for their collaborations. This work was supported by JST CREST and JSPS, KAKENHI Grants No. 15H02100, No. 23244062, No. 16K05459, No26887003.

## References

[1] P. P. Kapitza, *Soviet Phys. JETP* **21**, 588(1951).

[2] M. Bukov, L. D'Alessio, and A. Polkovnikov, *Advances in Physics*, **64**, 139(2015). (arXiv:1407.4803v6)

[3] M. Pont and M. Gavrila, *Phys. Rev. Lett.* **65**, 2362(1990).

[4] J. H. Eberly and K. C. Kulander, *Science* **262**, 1229(1993).

[5] R. W. Schoenlein, W. Z. Lin, J. G. Fujimoto, G. L. Eesley, *Phys. Rev. Lett.* **58**, 1680(1987).

[6] S. D. Brorson, A. Kazeroonian, J. S. Moodera, D. W. Face, T. K. Cheng, E. P. Ippen, M. S. Dresselhaus, and G. Dresselhaus, *Phys. Rev. Lett.* **64**, 2172(1990).

[7] J. Shah, *Ultrafast Spectroscopy of Semiconductors and Semiconductor Nanostructures*, 2nd edition (Springer, Berlin, 1999).

[8] R. D. Averitt, and A. J. Taylor, *J. Phys. Condens. Matter* **14**, R1357 (2002).

[9] R. Huber, F. Tauser, A. Brodschelm, M. Bichler, G. Abstreiter, and A. Leitenstorfer, *Nature* **414**, 286(2001)

[10] M. Hase, M. Kitajima, A. M. Constantinescu, and H. Petek, *Nature* **426**, 51 (2003).

[11] Y. Kawakami, T. Fukatsu, Y. Sakurai, H. Unno, H. Itoh, S. Iwai, T. Sasaki, K. Yamamoto, K. Yakushi, and K. Yonemitsu, *Phys. Rev. Lett.* **105**, 246402(2010).

[12] L. Waldecker, R. Bertoni, and R. Ernstorfer, and J. Vorberger, *Phys. Rev. X* **6**, 021003(2016).

[13] T. Ishikawa, Y. Sagae, Y. Naitoh, Y. Kawakami, H. Itoh, K. Yamamoto,

K. Yakushi, H. Kishida, T. Sasaki, S. Ishihara, Y. Tanaka, K. Yonemitsu and S. Iwai, *Nat. Commun.* **5**, 5528 (2014).

[14] D. H. Dunlap and V. M. Kenkre, *Phys. Rev.* B**34**, 3625 (1986).

[15] F. Grossmann, T. Dittrich, P. Jung, and P. Hänggi, *Phys. Rev. Lett.* **67**, 516 (1991).

[16] Y. Kayanuma: *Phys. Rev.* A **50,** 843 (1994).

[17] H. Aoki, N. Tsuji, M. Eckstein, M. Kollar, T. Oka, and P. Werner, *Rev. Mod. Phys.* **86**, 779 (2014).

[18] N. Tsuji, T. Oka, P. Werner, and H. Aoki, *Phys. Rev. Lett.* **106,** 236401 (2011).

[19] N. Tsuji, T. Oka, H. Aoki, and P. Werner, *Phys. Rev. B* **85,** 155124 (2012).

[20] K. Nishioka and K. Yonemitsu, *J. Phys. Soc. Jpn.* **83**, 024706(2014).

[21] H. Yanagiya, Y. Tanaka, and K. Yonemitsu, *J. Phys. Soc. Jpn.* **84**, 094705(2015).

[22] J. H. Mentink, K. Balzer, and M. Eckstein, *Nature commun.* **6**, 6708(2015).

[23] K. Yonemitsu, *J. Phys. Soc. Jpn.*, **86**, 024711(2017).

[24] A. Ono, H. Hashimoto, and S. Ishihara, *Phys. Rev.* B**94**, 115152(2016).

[25] A. Ono and S. Ishihara, *Phys. Rev. Lett.*, **119**, 207202(2017).

[26] K. Ido, T. Ohgoe, and M. Imada, *Sci. Adv.* **3**, e1700718(2017).

[27] K. Takasan, M. Nakagawa, and N. Kawakami, *Phys. Rev.* B**96**, 115120(2017).

[28] N. F. Mott, *Metal-Insulator Transition*, 2'nd edition(Tayler & Francis, 1990)

[29] M. Imada, A. Fujimori, and Y. Tokura, *Rev. Mod. Phys.* **70**, 1039(1998),

[30] "Special Topic: Organic Conductors" eds. S. Kagoshima, K. Kanoda, and T. Mori, *J. Phys. Soc. Jpn.* **75** (2006) 051001.

[31] J. Merino and R. H. McKenzie, *Phys. Rev. Lett.* **87**, 237002(2001).

[32] N. Takeshita, T. Sasagawa, T. Sugioka, Y. Tokura, and H. Takagi, *J. Phys. Soc. Jpn.* **73**, 1123(2004).

[33] N. Gomes, W. W. De Silva, T. Dutta, R. T. Clay, and S. Mazumdar, *Phys. Rev.* B**93**, 165110(2016).

[34] P. Monceau, F. Ya. Nad, and S. Brazovskii, *Phys. Rev. Lett.* **86**, 4080 (2001).

[35] J. van den Brink and D. I. Khomskii, *J. Phys. Condens. Matter* 20, 434217(2008).

[36] S. Ishihara, *J. Phys. Soc. Jpn.* **79**, 011010 (2010).

[37] E. J. W. Verwey, *Nature* **144**(1939) 327.

[38] "Special Topic: Photo-induced phase transition" eds. M. Kuwata-Gonokami, and S. Koshihara, *J. Phys. Soc. Jpn.* **75,** 011001-011008 (2006).

[39] K. Yonemitsu and K. Nasu, *Phys. Rep.* **465,** 1 (2008).

[40] D. N. Basov, R. D. Averitt, D. van der Marel, M. Dressel, and K. Haule, *Rev. Mod. Phys.* **83,** 471 (2011).

[41] D. Nicoletti and A. Cavalleri, *Advances in Optics and Photonics* **8**, 401 (2016).

[42] C. Giannetti, M. Capone, D. Fausti, M. Fabrizio, F. Parmigiani, and D. Mihailovic, *Advances in Physics*, **65**, 58(2016).

[43] K. Onda, S. Ogihara, K. Yonemitsu, N. Maeshima, T. Ishikawa, Y. Okimoto, X. Shao, Y. Nakano, H. Yamochi, G. Saito, and S. Koshihara, *Phys. Rev. Lett.*, **101**, 067403(2008).

[44] L. Guérin, J. Hébert, M. B-L. Cointe, S. Adachi, S. Koshihara, H. Cailleau, and E. Collet, *Phys. Rev. Lett.* **105**, 246101 (2010).

[45] M. Servol, N. Moisan, E. Collet, H. Cailleau, W. Kaszub, L. Toupet, D. Boschetto, T. Ishikawa, A. Moréac, S. Koshihara, M. Maesato, M. Uruichi, X. Shao, Y. Nakano, H. Yamochi, G. Saito, and M. Lorenc, *Phys. Rev.* B**92**, 024304(2015).

[46] K. Itoh, H. Itoh, M. Naka, S. Saito, I. Hosako, N. Yoneyama, S. Ishihara, T. Sasaki, and S. Iwai, *Phys. Rev. Lett.* **110**, 106401 (2013).

[47] H. Matsuzaki, M. Iwata, T. Miyamoto, T. Terashige, K. Iwano, S. Takaishi, M. Takamura, S. Kumagai, M. Yamashita, R. Takahashi, Y. Wakabayashi, and H. Okamoto, *Phys. Rev. Lett.* **113**, 096403 (2014).

[48] L. Rettig, R. Cortés, J.-H. Chu, I. R. Fisher, F. Schmitt, R. G. Moore, Z.-X. Shen, P. S. Kirchmann, M. Wolf, and U. Bovensiepen, *Nat. Commun.* **7**, 10459 (2016).

[49] A. Singer, S. K. K. Patel, R. Kukreja, V. Uhlı́řr, J. Wingert, S. Festersen, D. Zhu, J. M. Glownia, H. T. Lemke, S. Nelson, M. Kozina, K. Rossnagel, M. Bauer, B. M.Murphy, O. M.Magnussen, E. E. Fullerton, and O. G. Shpyrko, *Phys. Rev. Lett.* **117** , 056401 (2016).

[50] K. W. Kim, A. Pashkin, H. Schäfer, M. Beyer, M. Porer, T. Wolf, C. Bernhard, J. Demsar, R. Huber, and A. Leitenstorfer, *Nat. Mater.* **11**, 497(2012).

[51] D. Fausti, R. I. Tobey, N. Dean, S. Kaiser, A. Dienst, M. C. Hoffmann, S. Pyon, T. Takayama, H. Takagi, and A. Cavalleri, *Science* **331**, 189 (2011).

[52] S. Kaiser, C. R. Hunt, D. Nicoletti, W. Hu, I. Gierz, H. Y. Liu, M. Le Tacon, T. Loew, D. Haug, B. Keimer, and A. Cavalleri, *Phys. Rev.* B**89**, 184516(2014).

[53] R. Matsunaga, N. Tsuji, H. Fujita, A. Sugioka, K. Makise, Y. Uzawa, H. Terai, Z. Wang, H. Aoki, and R. Shimano, *Science* 345, 1145(2014).

[54] Y. Kawakami, Y. Yoneyama, T. Amano, H. Itoh, K. Yamamoto, Y. Nakamura, H. Kishida, T. Sasaki, S. Ishihara, Y. Tanaka, K. Yonemitsu, and S. Iwai, Phys. Rev. B**95**, 201105(R)(2017).

[55] Y. Naitoh, Y. Kawakami, T. Ishikawa, Y. Sagae, H. Itoh, K. Yamamoto, T. Sasaki, M. Dressel, S. Ishihara, Y. Tanaka, K. Yonemitsu, and S. Iwai, *Phys. Rev.* B **93**,165126 (2016).

[56] R. L. Fork, C. H. B. Cruz, P. C. Becker and C. V. Shank, *Opt. Lett.* **12** , 483(1987).

[57] M. Nisoli, S. De Silvestri, O. Svelto, *Appl. Phys. Lett.* **68,** 2793(1996).

[58] B. E. Schmidt, A. D. Shiner, P. Lassonde, J-C. Kieffer, P. B. Corkum, D. M. Villeneuve, F. Légaré, *Opt. Express* **19**, 6858(2011).

[59] V. Cardin, N. Thiré, S. Beaulieu, V. Wanie, F. Légaré and B. E. Schmidt, *Appl. Phys. Lett.* **107** 181101(2015).

[60] D. Brida, G. Cirmi, C. Manzoni, S. Bonora, P. Villoresi, S. De Silvestri and G. Cerullo, *Opt. Lett.* **33**, 741 (2008).

[61] K. Bender, I. Hennig, D. Schweitzer, K. Dietz, H. Endres, and H. J. Keller, *Mol. Cryst., Liq. Cryst.* **108**, 359(1984).

[62] A. Pashkin, M. Dressel, and C. A. Kuntscher, *Phys. Rev.* B**74**, 165118(2006).

[63] H. Seo, *J. Phys. Soc. Jpn.*, **69**, 805(2000).

[64] Y. Takano, K. Hiraki, H. M. Yamamoto, T. Nakamura, and T. Takahashi, *Synth. Met.* **120**, 1081(2001).

[65] R. Wojciechowski, K. Yamamoto, K. Yakushi, M. Inokuchi, and A. Kawamoto, *Phys. Rev.* B **67**, 224105(2003).

[66] T. Kakiuchi, Y. Wakabayashi, H. Sawa, T. Takahashi, and T. Nakamura, *J. Phys. Soc. Jpn.*, 76, 113702(2007).

[67] Y. Tanaka and K. Yonemitsu, *J. Phys. Soc. Jpn.* **77**, 034708(2008).

[68] T. J. Emge, P. C. W. Leung, M. A. Beno, H. H. Wang, J. M. Williams, M-H. Whangbo, and M. Evain, *Mol. Cryst. Liq. Cryst.* 138, 393 (1986).

[69] T. Mori, H. Mori, and S. Tanaka, *Bull. Chem. Soc. Jpn.* **72**, 179(1999).

[70] S. Iwai, K. Yamamoto, A. Kashiwazaki, F. Hiramatsu, H. Nakaya, Y. Kawakami, K. Yakushi, H. Okamoto, H. Mori, and Y. Nishio, *Phys. Rev. Lett.* **98**, 097402 (2007).

[71] K. Yamamoto, S. Iwai, S. Boyko, A. Kashiwazaki, F. Hiramatsu, C. Okabe, N. Nishi, and K. Yakushi, *J. Phys. Soc. Jpn.* **77**, 074709(2008).

[72] S. Iwai, K. Yamamoto, F. Hiramatsu, H. Nakaya, Y. Kawakami, and K. Yakushi, *Phys. Rev.* B **77**, 125131(2008).

[73] H. Nakaya, K. Itoh, Y. Takahashi, H. Itoh, S. Iwai, S. Saito, K. Yamamoto, and K. Yakushi,, *Phys. Rev.* B**81**,155111(2010).

[74] S. Iwai, *J. Lumin.* **131**, 409(2011).

[75] H. Itoh, K. Itoh, K. Goto, K. Yamamoto, K. Yakushi, and S. Iwai, *Appl. Phys. Lett.* **104**, 173302(2014).

[76] S. Iwai, *Crystals* **2**, 590(2012).

[77] S. Miyashita, Y. Tanaka, S. Iwai, and K. Yonemitsu, *J. Phys. Soc. Jpn.* **79**, 034708(2010).

[78] H. Gomi, A. Takahashi, T. Tatsumi, S. Kobayashi, K. Miyamoto, J. D. Lee, and M. Aihara, *J. Phys. Soc. Jpn.*, **80**, 034709(2011).

[79] K. Iwano, *Phys. Rev.* B**91**, 115108(2015).

[80] H. Hashimoto, H. Matsueda, H. Seo, and S. Ishihara, *J. Phys. Soc. Jpn.* **84**, 113702(2015).

[81] T. Sugano, K. Yamada, G. Saito, and M. Kinoshita, Solid State commun. 55, 137 (1985).

[82] S. Sugai, and G. Saito, Solid State Commun. 58, 759(1986).

[83] V. Zelezny, J. Petzelt, R. Swietlik, B.P. Gorshunov, A.A. Volkov, G.V. Kozlov, D. Schweitzer, and H.J. Keller, *J. Phys. France* **51**, 869(1990).

[84] M. Dressel, G. Grüner, J. P. Pouget, A. Breining, and D. Schweitzer, *J. Phys. I France* **4,** 579(1994).

[85] Y. Yue, K. Yamamoto, M. Uruichi, C. Nakano, K. Yakushi, S. Yamada, T. Hiejima, and A. Kawamoto, *Phys. Rev. B* **82,** 075134 (2010).

[86] K. Yakushi, *Crystals* 2, 1291 (2012).

[87] C. S. Jacobsen, D. B. Tanner, and K. Bechgaard, *Phys. Rev.* B**28**, 7019 (1983).

[88] L. Balicas, K. Behnia, W. Kang, E. Canadell, P. Auban-Senzier, D. Jerome, M. Ribault, and J. M. Fabre, *J. Phys. I France*, **4**, 1539 (1994).

[89] M. Dressel, M. Dumm, T. Knoblauch, and M. Masino, *Crystals* **2**, 528 (2012).

[90] M. Dressel and G. Grüner, Electrodynamics of Solids, (Cambridge University Press, Cambridge, 2002 )

[91] Y. Naitoh, Y. Kawakami, T. Ishikawa, Y. Sagae, H. Itoh, K. Yamamoto, T. Sasaki, M. Dressel, S. Ishihara, Y. Tanaka, K. Yonemitsu, and S. Iwai, Supplemental material at http://link.aps.org/supplemental/10.1103/PhysRevB.93.165126

[92] J. Favand and F. Mila, *Phys. Rev.* B**54**, 10425(1996).

[93] Y. Kawakami, T. Amano, Y. Yoneyama, Y. Akamine, H. Itoh, G. Kawaguchi, H. M. Yamamoto, H. Kishida, K. Itoh, T. Sasaki, S. Ishihara, Y. Tanaka, K. Yonemitsu, and S. Iwai, *Nat. Photon.* **12**, 474(2018).

Figure captions

**Figure 1**

(a) Kapitza pendulum[1], (b) Schematic illustration of dynamical localization for tight-binding model [13]. (c) Left side: Schematic illustration of Peierls-phase $\phi(t)$ for vector potential $A(t) = -c\int_0^t E(t')dt'$, where $E(t)$ represents an electric field of light. Right side: zeroth order Bessel function $J_0$ which is proportional to effective transfer integral $t_{\text{eff}}$ as a function of $eE_0 a/\hbar\omega$ ($E_0$: amplitude of light-field, $a$: inter-site distance, $\omega$: angular frequency of light).

**Figure 2**

(a) Schematic illustration of 7-fs light source with photos of Kr-filled chamber and active-mirror-compressor, and interference pattern of 2f-3f interferometer for carrier-envelope-phase (CEP) measurement. (b) Schematic illustration of active mirror compressor. (c)(d) Spectrum(c) and auto-correlation profile(d) after chamber (before compression). The solid lines and circles respectively show experimental and calculated results. The dashed line shows the spectrum of the idler pulse from the optical parametric amplifier. (e) Auto-correlation profile after compression (pulse width is 6 fs).

**Figure 3**

Light-induced short-range CO in α-(BEDT-TTF)$_2$I$_3$ and its detection by pump-probe reflectivity measurement (section 4). The light intensity after a pair of wire-grid CaF$_2$ polarizers is also shown as a function of angles $\theta_1$ and $\theta_2$ [54].

Figure 4

(a) Crystal structure of α-(BEDT-TTF)$_2$I$_3$. (b) CO insulator-to-metal transition (red arrow, melting) and metal-to-CO insulator transition (blue arrow, freezing) for both thermal(equilibrium) and optical(non-equilibrium) transitions in α-(BEDT-TTF)$_2$I$_3$ [13].

Figure 5

(a) $\sigma$ spectra of α-(BEDT-TTF)$_2$I$_3$ at 40 K (CO: dashed blue curve) and at 140 K (metal: solid red curve). The spectrum of the 7 fs pulse is indicated by the orange curve[13]. (b) $R$ spectra at 40 K (CO) and at 140 K (metal)[13]. (c) The spectra for three temperature differentials: [$R$(40K) - $R$(140K)]/$R$(140K) (solid curve with blue shading), [$R$(190K) - $R$(140K)]/$R$(140 K) [dashed curve (x3) with red shading], and [$R$(170K) - $R$(140K)]/$R$(140 K) [dashed-dotted curve (x3)][13]. (d) Reflectivities measured at 0.09 eV and 0.64 eV [indicated by the blue arrows on the spectrum in Fig. 5(b)], plotted as a function of temperature[53].

Figure 6

(a) $\Delta R/R$ spectra for $I_{ex}$ = 0.8 and 0.12 mJ/cm$^2$ at $t_d$ = 30 fs (closed blue circles for $I_{ex}$ =0.8 mJ/cm$^2$, open blue circles for $I_{ex}$ =0.12 mJ/cm$^2$) and 300 fs (closed black circles for $I_{ex}$ = 0.8 mJ/cm$^2$) after excitation by a 7 fs pulse. $\Delta R/R$ at $t_d$ =

300 fs after a 100-fs pulse excitation for $I_{ex}$ = 0.8 mJ/cm$^2$ is shown as the crosses[13]. **(b)** Temporal change of the $\Delta R/R$ spectrum, plotted as a 2-dimensional (probe energy–delay time) map. Positive and negative $\Delta R/R$ are shown by the blue and red shadings, respectively[13].

**Figure. 7**

**(a)** Time evolution of $\Delta R/R$ observed at 0.64 eV for $I_{ex}$ = 0.8 mJ/cm$^2$ (solid curve) and 0.12 mJ/cm$^2$ (dashed-dotted curve). The cross correlation between the pump and the probe pulses is also indicated by the orange shading. The dashed blue curve shows the positive component of $\Delta R/R$, reflecting the photoinduced charge localization, which was estimated assuming that the negative component grows exponentially (dashed red curve)[13]. **(b)** Oscillating component of the time profile. The time-resolved spectra of the oscillating component obtained by wavelet (WL) analysis (blue curve for 0–40 fs, red curve for 80–120 fs) are shown in the inset. The optical conductivity spectra at 10 K (CO) and 140 K (metal) are shown by the black curves in the inset[13].

**Figure 8**

The hole densities ($\rho_H$) at the A and A' molecules in Fig. 4(b) as a function of the normalized temperature $T/T_{CO}$, calculated using the Hartree–Fock approximation for an extended Hubbard model[67] for $t_0$, $0.95t_0$, $0.9t_0$, and $0.85t_0$ (circles, squares, triangles, and inverted triangles, respectively)[13].

**Figure 9**

(a)(b) Lattice structure of α-(BEDT-TTF)$_2$I$_3$ in CO phase(a) and that in metallic phase(b)[66, 67]. Sites(=BEDT-TTF molecules) A, A', B and C are crystallographically non-equivalent. $b_1(b_1')$-$b_4(b_4')$, $a_1(a_1')$, $a_2$ and $a_3$ are the bonds between those sites [54].

**Figure 10**

(a)(b) $R$ spectra (0.08-0.8 eV (a), 0.4-0.8 eV(b)) measured for various polarizations ($\theta$=0º, 20º, 40º, 60º, 90º) at 138 K[54]. (c) $\Delta R/R$ spectra at the $t_d$ of 30 fs are shown for $\theta_{pr}$=0º (//b) (red-filled circles) and $\theta_{pr}$=90º (//a) (blue squares) for $\theta_{pu}$=0º (//b). The excitation intensity $I_{ex}$ was 0.9 mJ/cm² (9.8 MV/cm). The spectral differences between 138 K (metal) and 50 K (insulator) $(\Delta R/R)_{MI} = \{R(50K) - R(138K)\}/R(138K)$ are also shown for $\theta$=0º, 20º, 40º, 60º, 90º[54].

**Figure 11**

$\Delta R/R$ at 0.65 eV, $t_d$=30 fs as a function of $\theta_{pu}$ for $\theta_{pr}$=0º(red filled circles), 23º(green rectangles), -23º(orange triangles), 90º(blue triangles)[54].

**Figure 12**

Field induced changes in spatially and temporally averaged correlation functions as indexes of the short-range CO[54]. Double occupancy (a)

$\langle\langle n_{i\uparrow}n_{i\downarrow}\rangle\rangle$. Nearest-neighbor density-density correlations (b) $\langle\langle n_i n_j\rangle\rangle_{a2,a3}$ for the $a_2$ and $a_3$ bonds and (c) $\langle\langle n_i n_j\rangle\rangle_b$ for the $b$ bonds. (d) $\theta_{pu}$ dependence of the averaged nearest-neighbor density-density correlation for the $a_2$ and $a_3$ bonds $\langle\langle n_i n_j\rangle\rangle_{a2,a3}$ for $eaF/\eta\omega=0$ (black crosses), 0.14 (blue rhombuses), 0.20 (green triangles), and 0.27 (red squares)[54].

# Figure 13

Crystal structure of $(TMTTF)_2AsF_6$.

# Figure 14

(a) $\sigma$ spectra of $(TMTTF)_2AsF_6$ at 25 K and 150 K. The spectrum of the 7-fs pulse (orange shaded area) is also shown[55]. (b) $R$ spectra at 25 K and 150 K with Lorentz analysis (solid lines). Inset shows the temperature dependence of $\Delta\gamma/\gamma$ [55].

# Figure 15

(a) Time evolution of $\Delta R/R$ at $t_d < 4$ ps measured by 100-fs pulse[55]. (b) $\Delta R/R$ spectrum at $t_d = 0.1$ ps with Lorentz analysis (solid line)[55]. (c) Spectral change calculated with only decreasing $\omega_p$ (1.8%) (blue) and that with only increasing $\gamma$ (12%)(red). A 11% increase in the width of the vibrational peak at 0.165 eV is taken into account[55].

# Figure 16

(a) Time evolution of $\Delta R/R$ at $t_d < 150$ fs measured by a 7-fs pulse[55]. (b) Time evolutions of $\Delta R/R$ measured at i) 0.85, ii) 0.73, and iii) 0.62 eV[55].

**Figure 17**

$\Delta R/R$ spectra at various time delays $t_d =$ 12 fs**(a)**–80 fs**(f)** are shown as the circles. The blue-dashed, red-dashed, and orange curves indicate the calculated spectral change using the Lorentz model (orange), the spectral change calculated with only $-\Delta\omega_p/\omega_p$ (blue-dashed) and that with only $\Delta\gamma/\gamma$ (red-dashed). The arrows indicate $\hbar\omega_p$ [55].

**Figure 18**

**(a)** Schematic illustration of response function $\chi(t)$ for oscillating charges with frequency $\omega_p$ and phase shift at $t_d$=40 fs **(b)** Spectrogram calculated by windowed Fourier transform. **(c)** Time-resolved spectral response function $\chi(\omega)$ at $t_d$=20 fs (black line) and at 40 fs (red line) in spectrogram **(b)**.

**Figure 19**

**(a)** Time profiles of $-\Delta\omega_p/\omega_p$ (blue dots) and $\Delta\gamma/\gamma$ (red dots) obtained using Lorentz analysis (shown by the blue-dashed and red-dashed curves in Fig. 17)[55]. **(b)** Time profile of the spectral area of the dip (green shade in Fig. 17)[55].

(a)

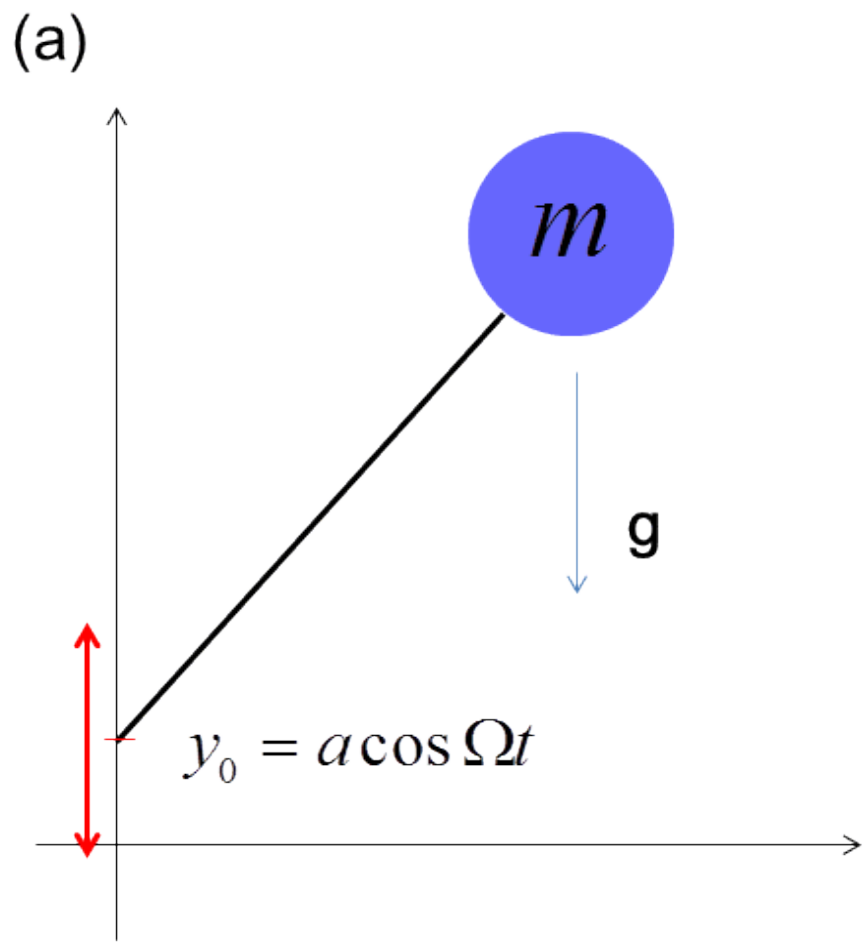

Kawakami et al.
Fig. 1(a)

(b)

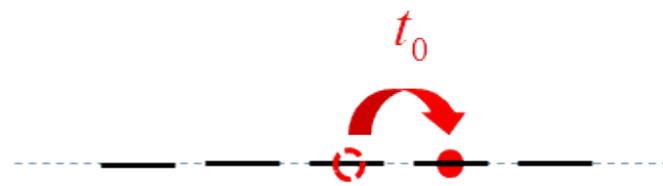

Tight binding approximation

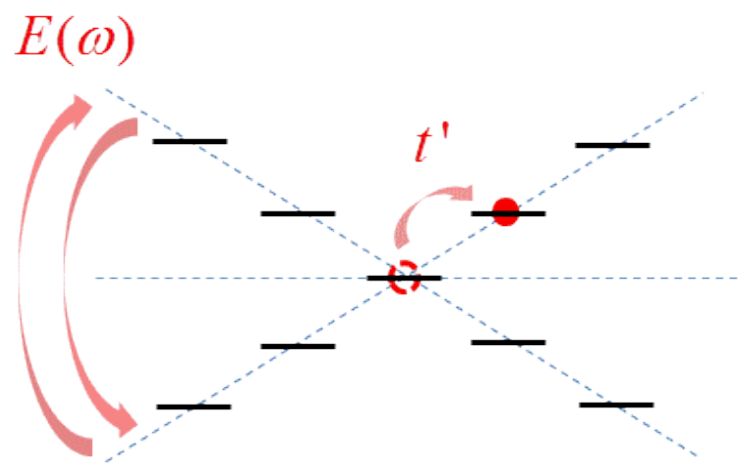

Kawakami et al.
Fig. 1(b)

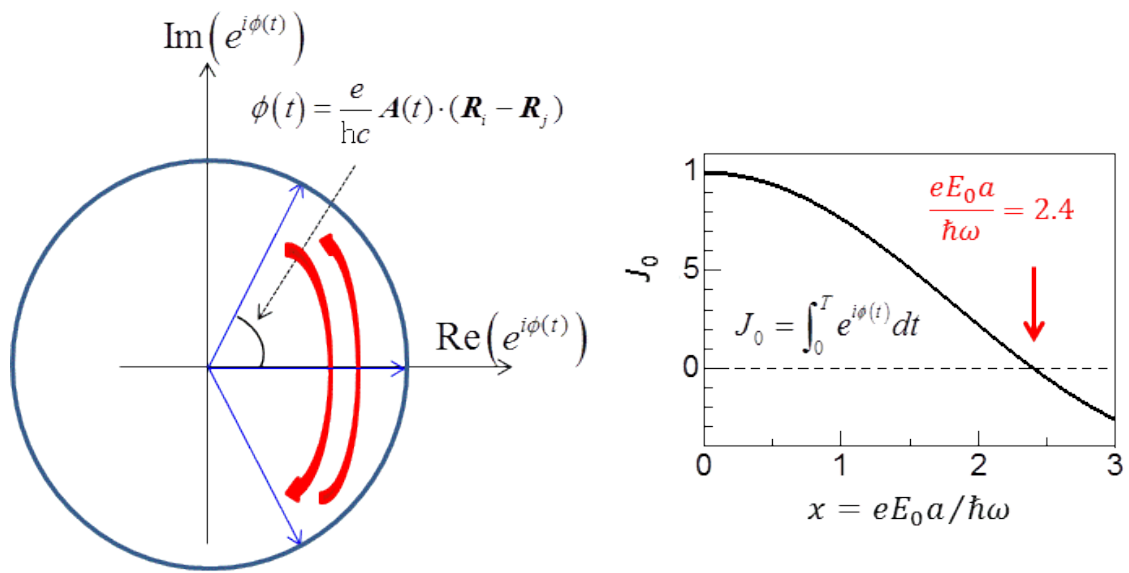

Kawakami et al.
Fig. 1(c)

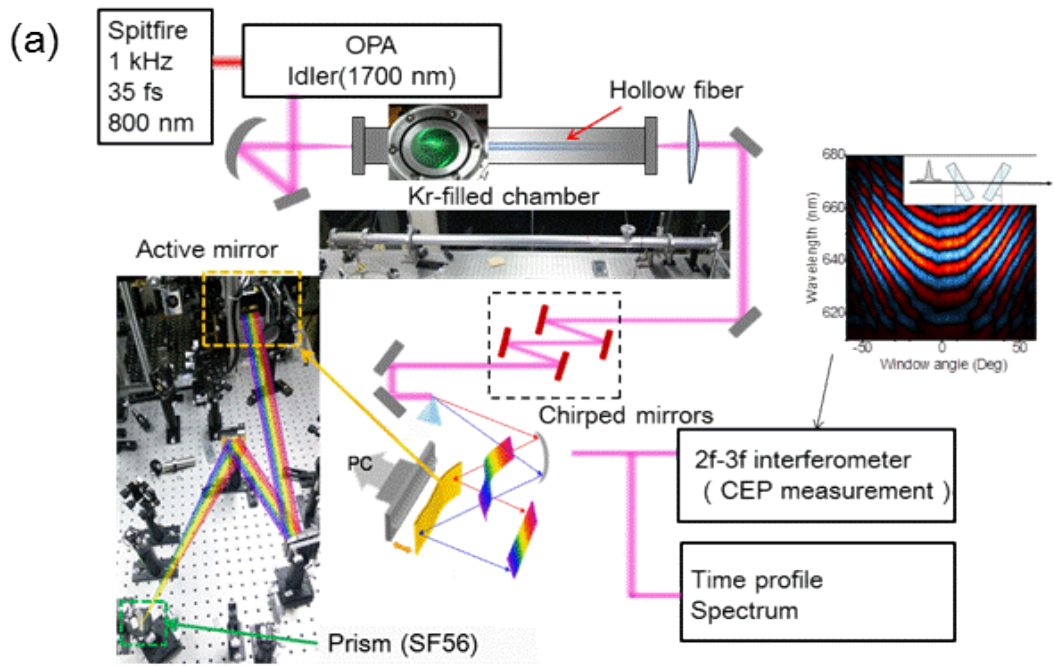

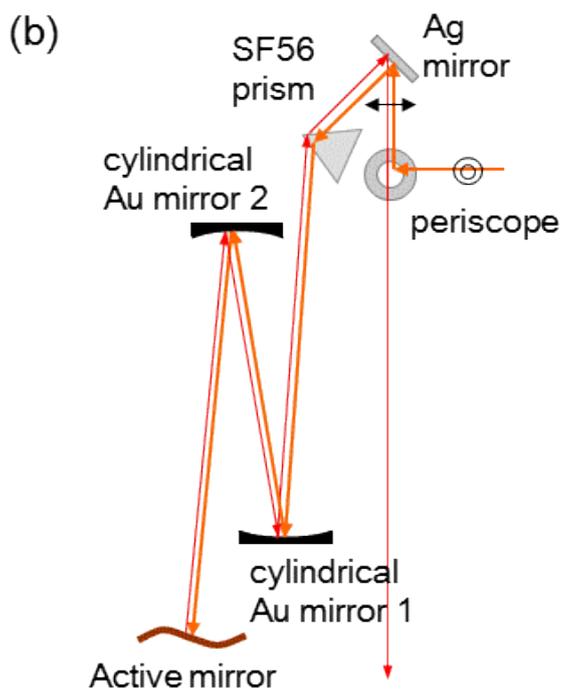
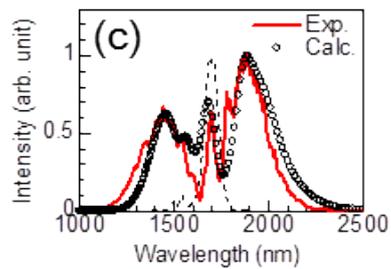
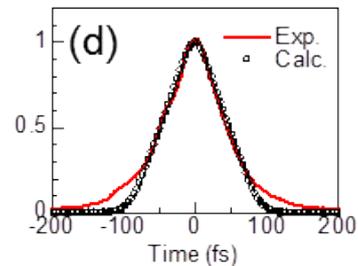
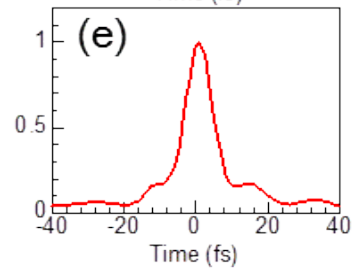

Kawakami et al. Fig. 2

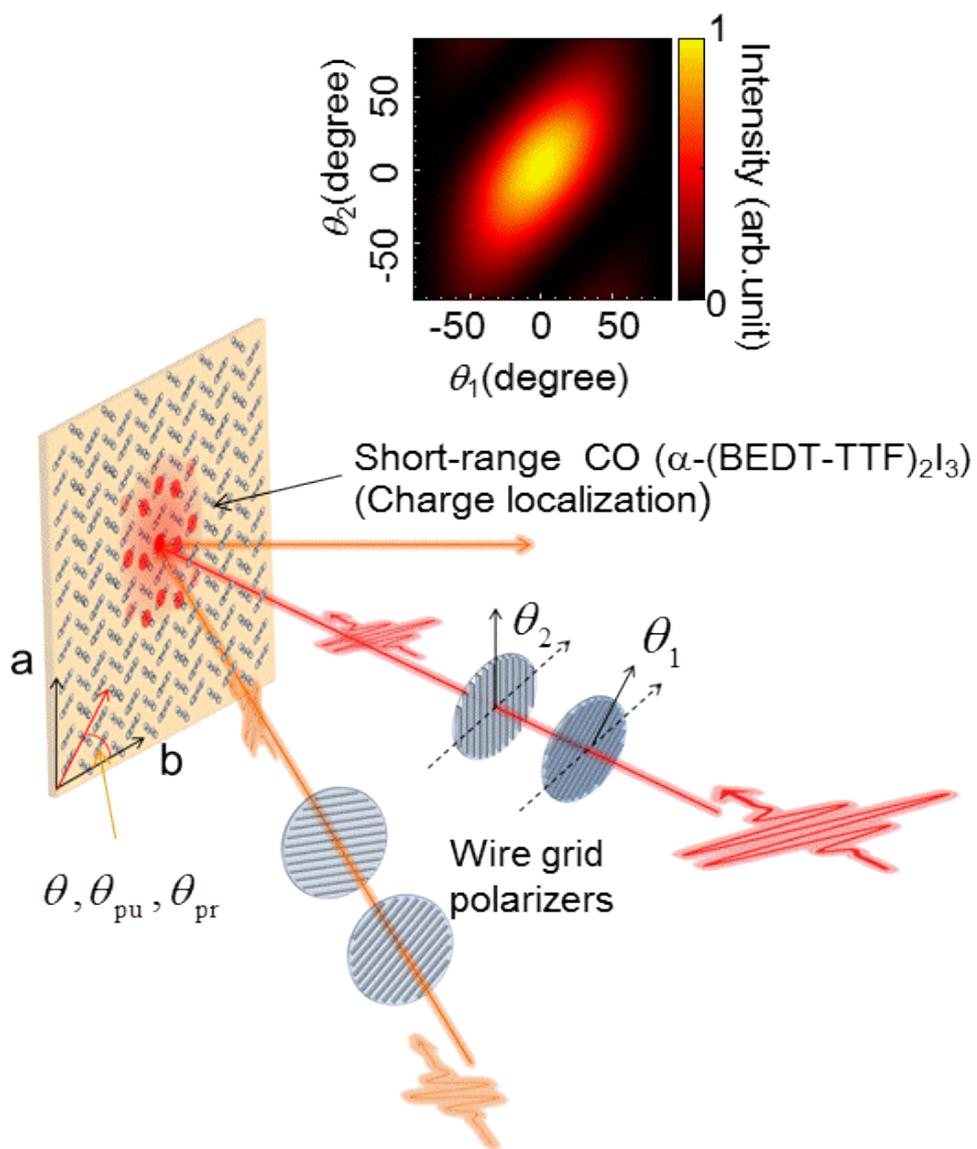

Kawakami et al.  Fig. 3

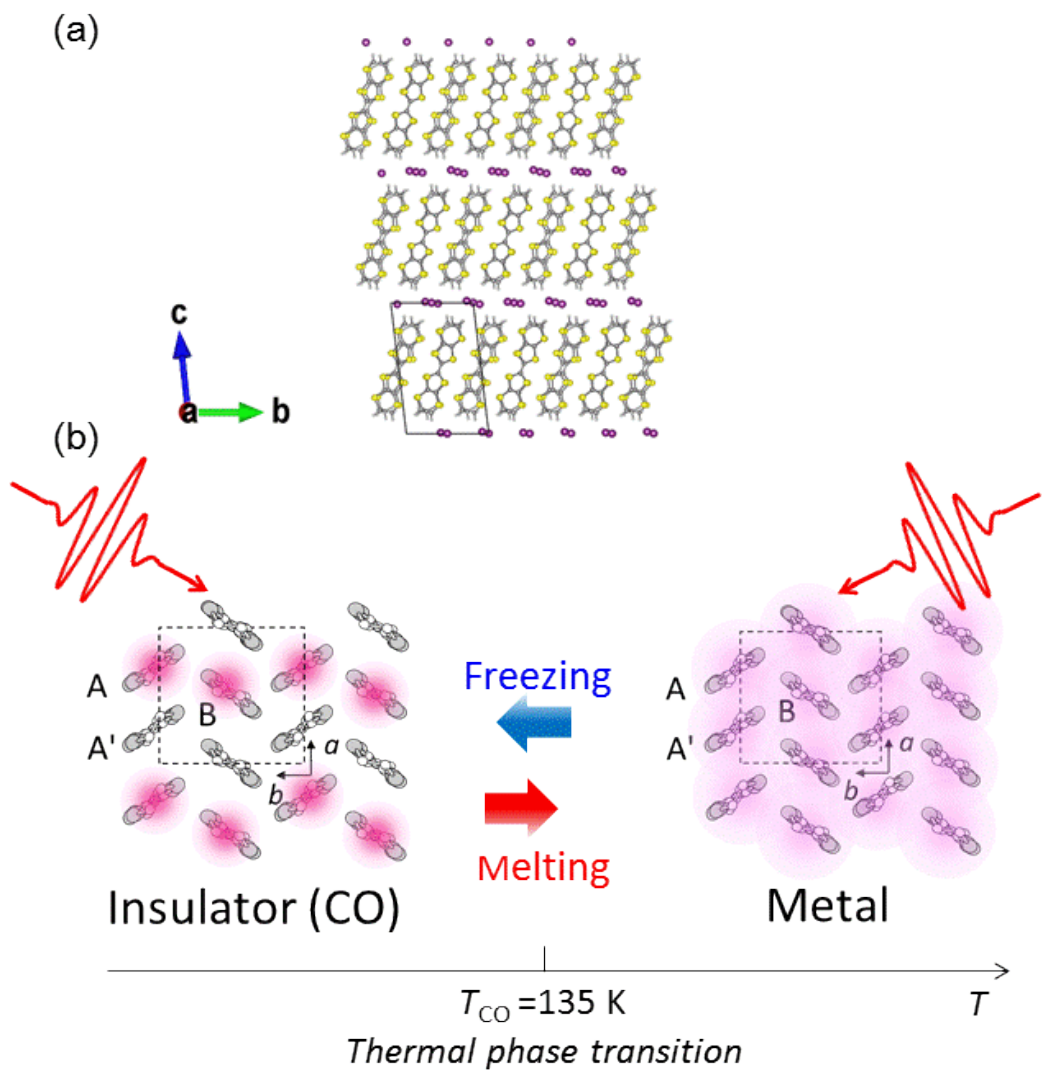

Kawakami et al.
Fig. 4

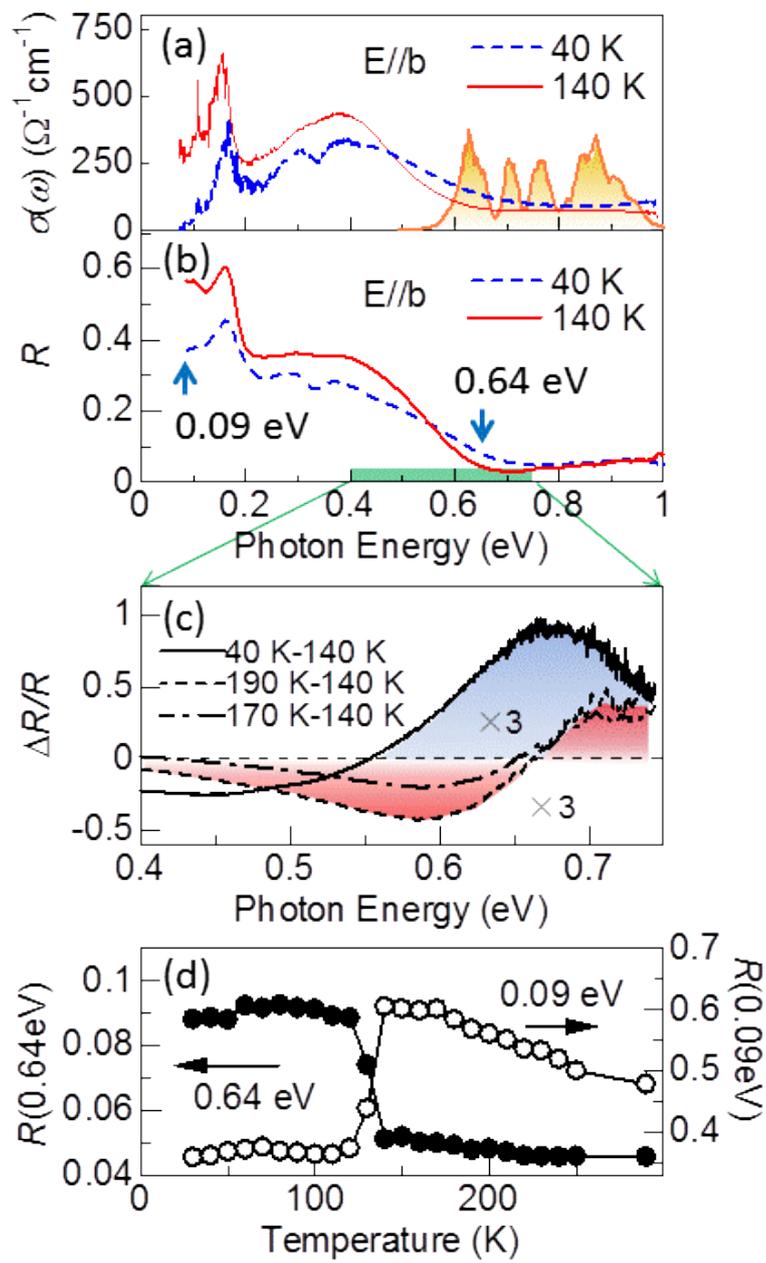

Kawakami et al.
Fig. 5

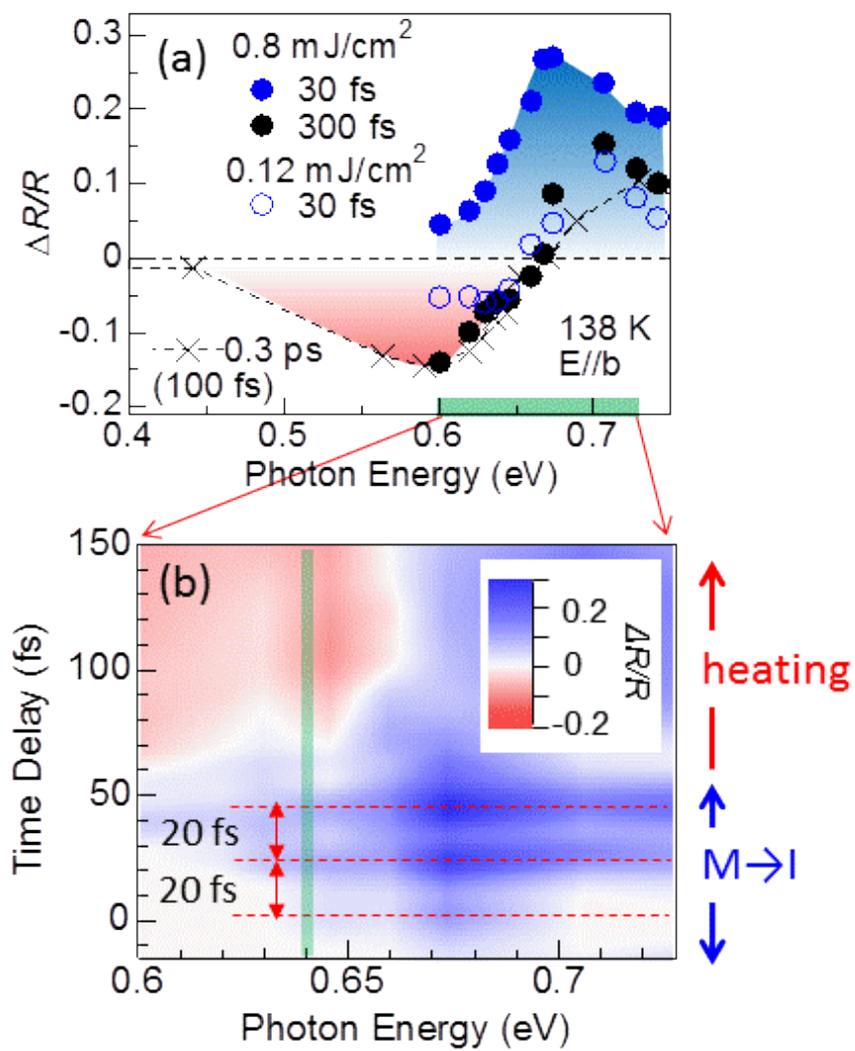

Kawakami et al., Fig. 6

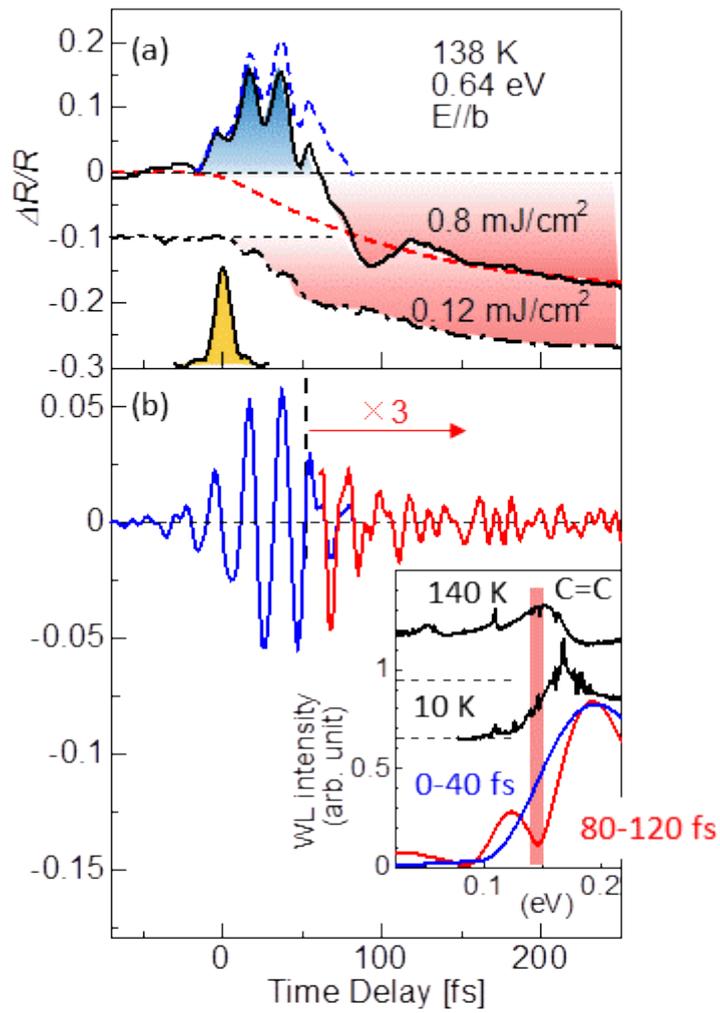

Kawakami et al.
Fig. 7

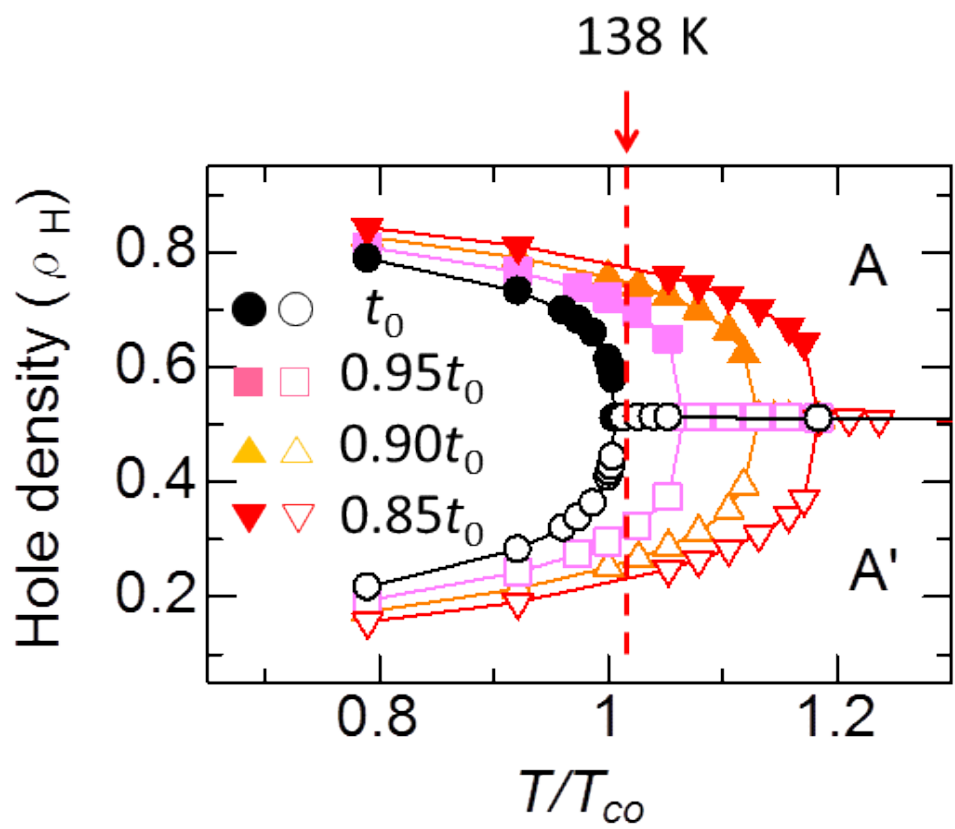

Kawakami et al.
Fig. 8

(a) 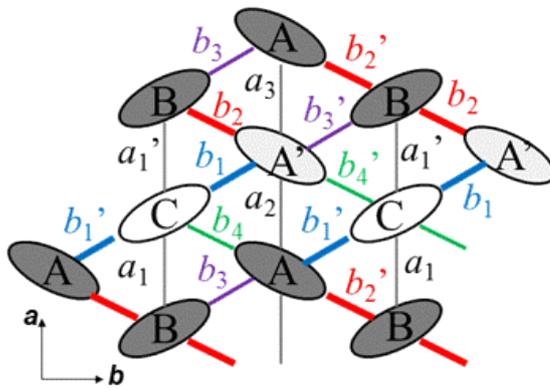

(b) 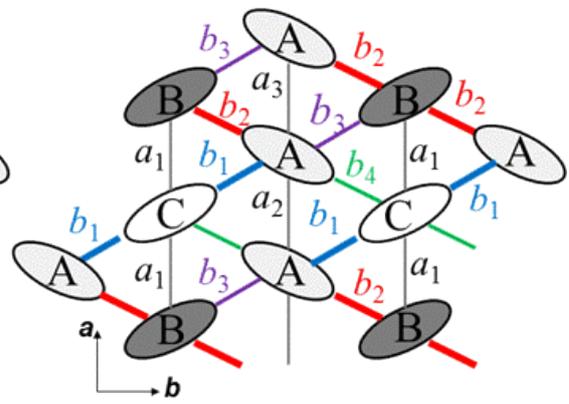

Kawakami et al.
Fig. 9

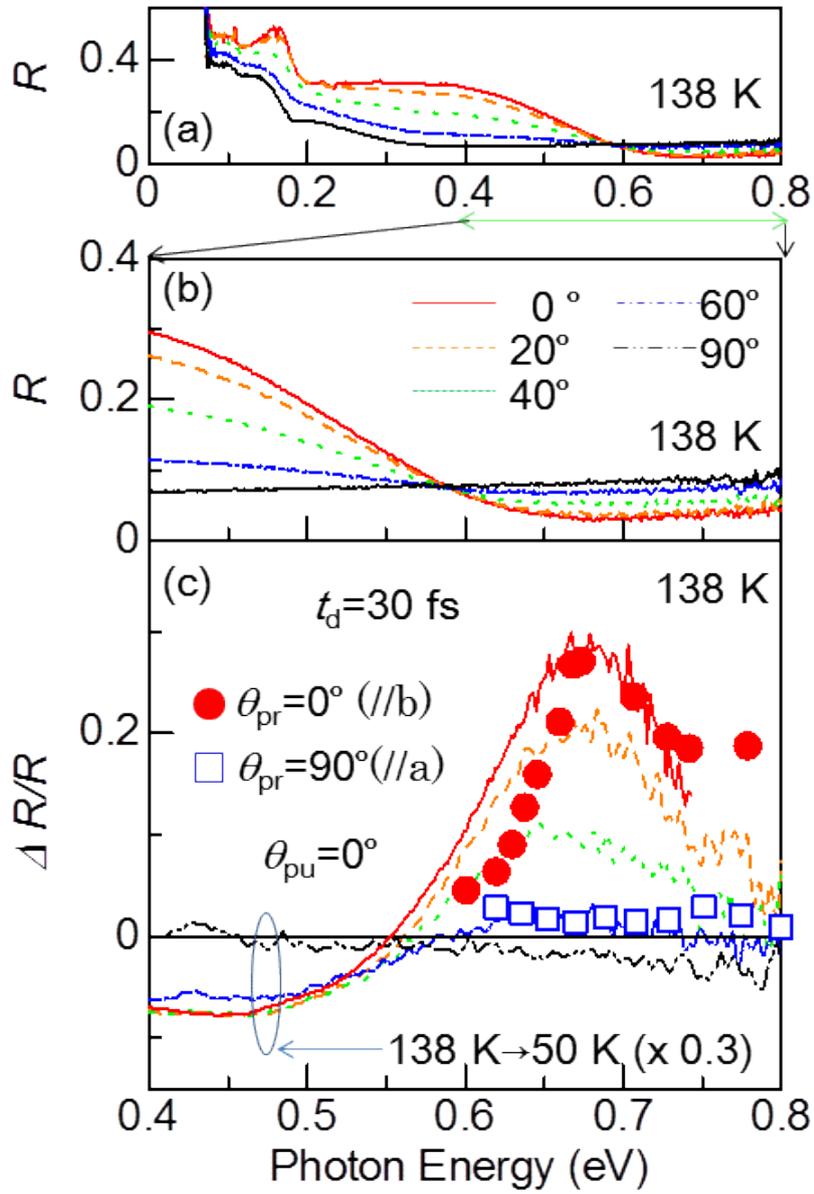

Kawakami et al.
Fig. 10

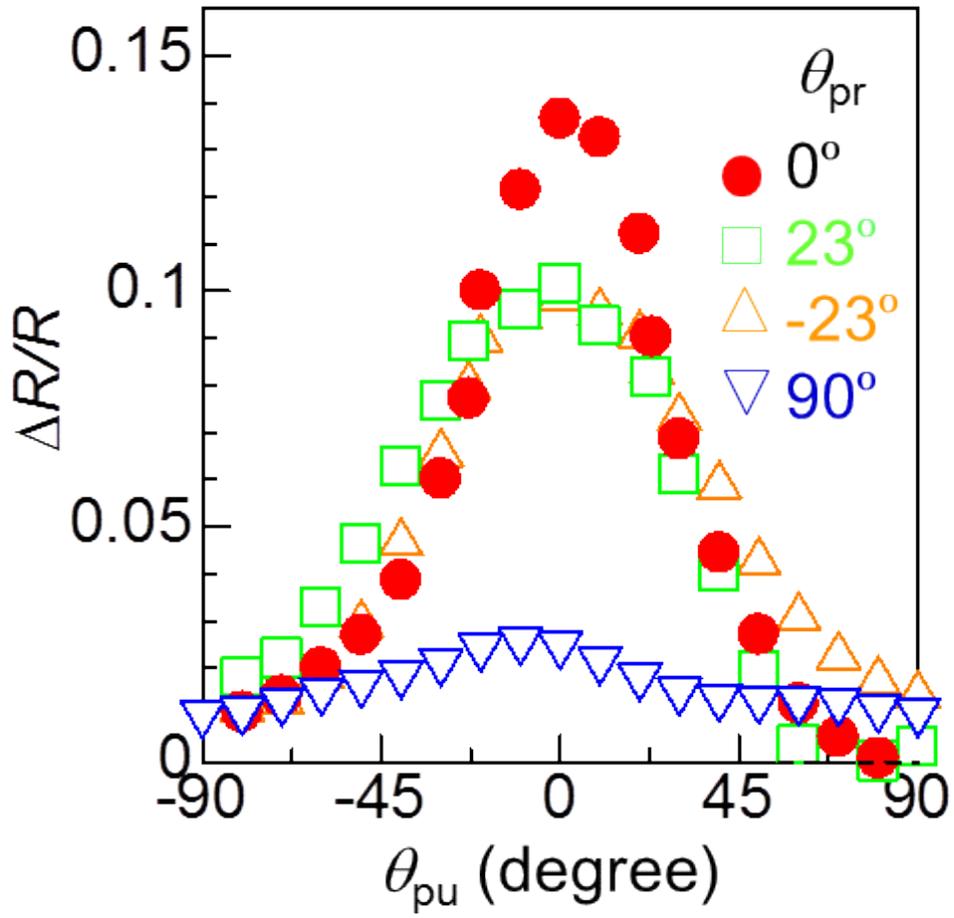

Kawakami et al.
Fig. 11

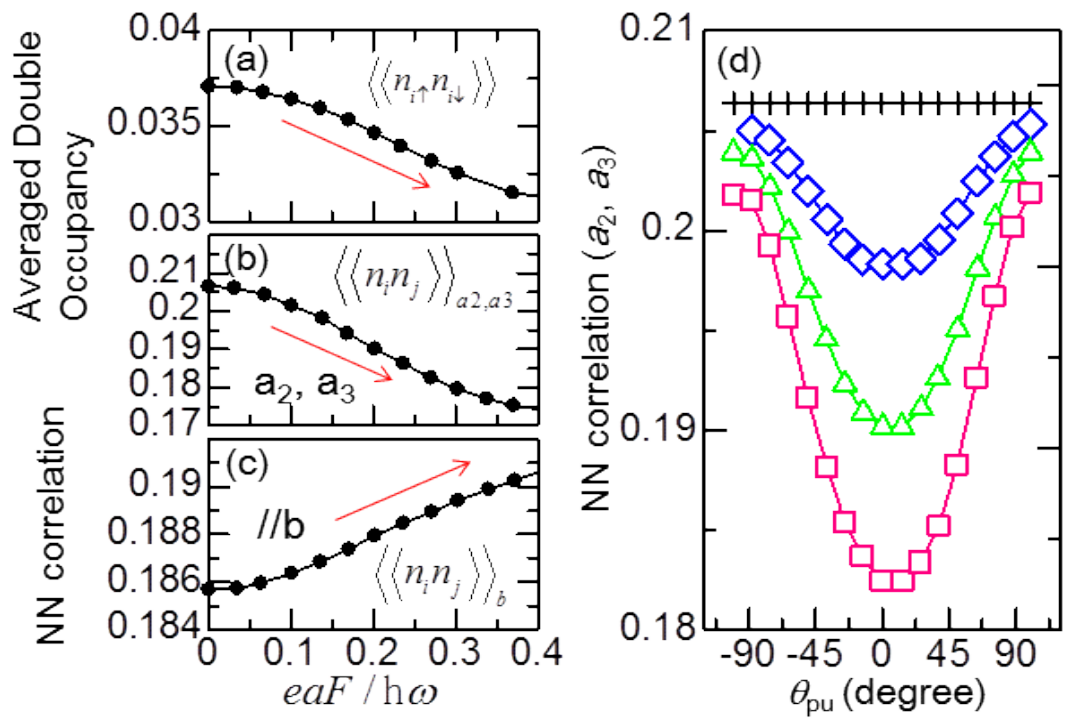

Kawakami et al. Fig. 12

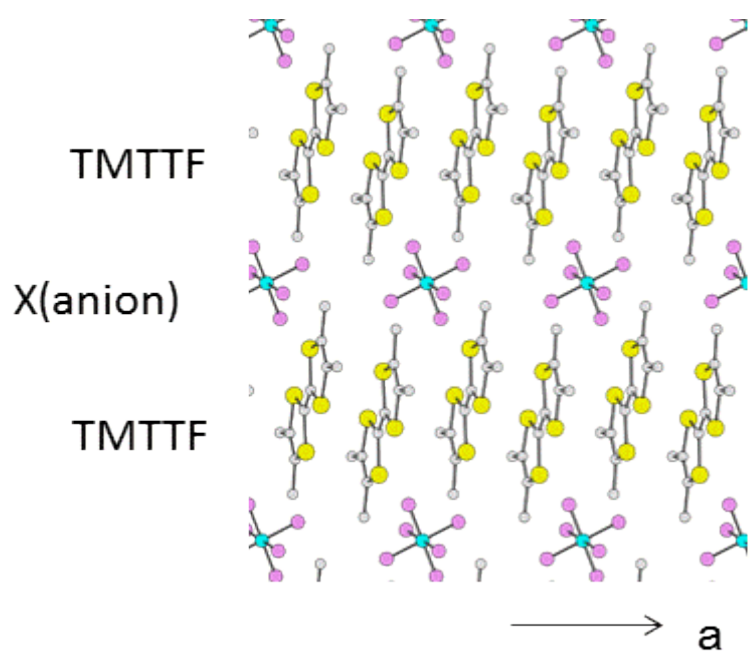

Kawakami et al. Fig.13

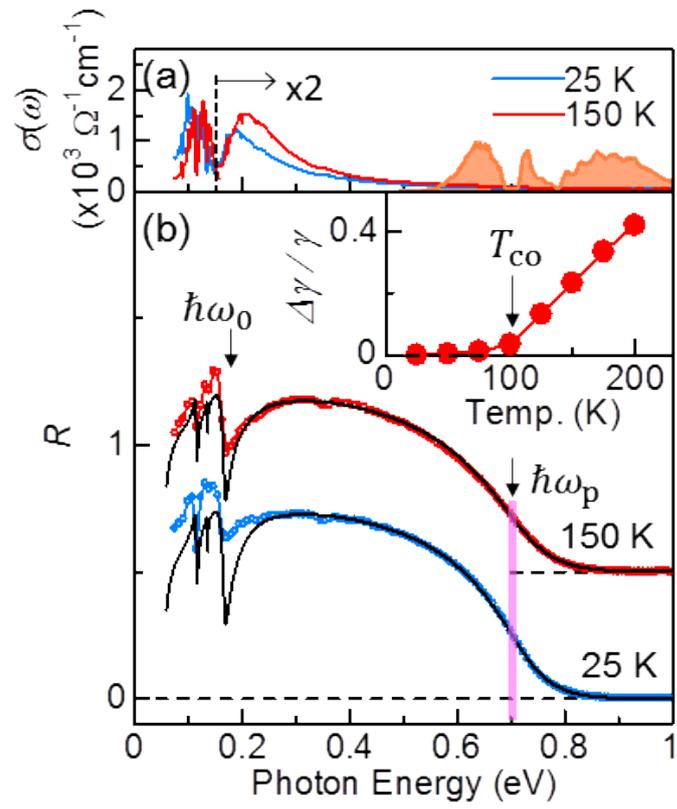

Kawakami et al. Fig. 14

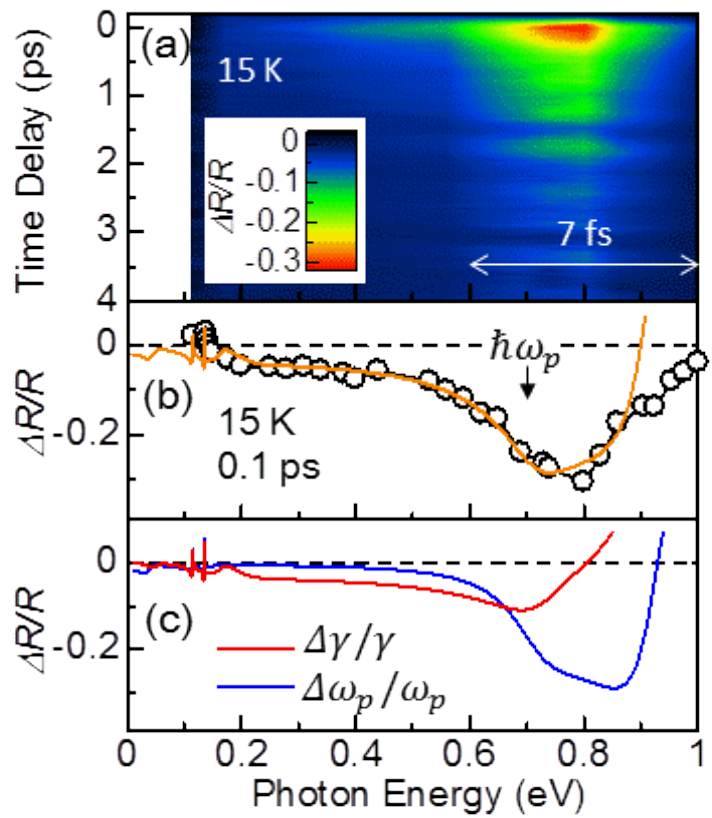

Kawakami et al. Fig. 15

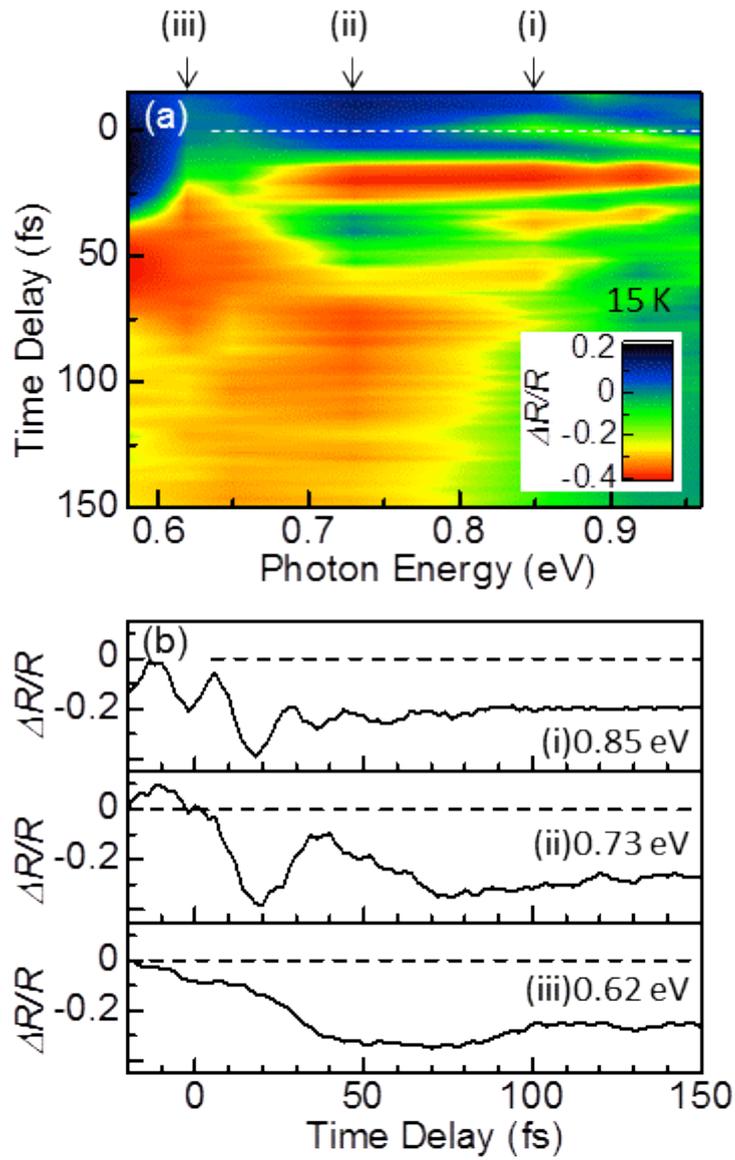

Kawakami et al., Fig. 16

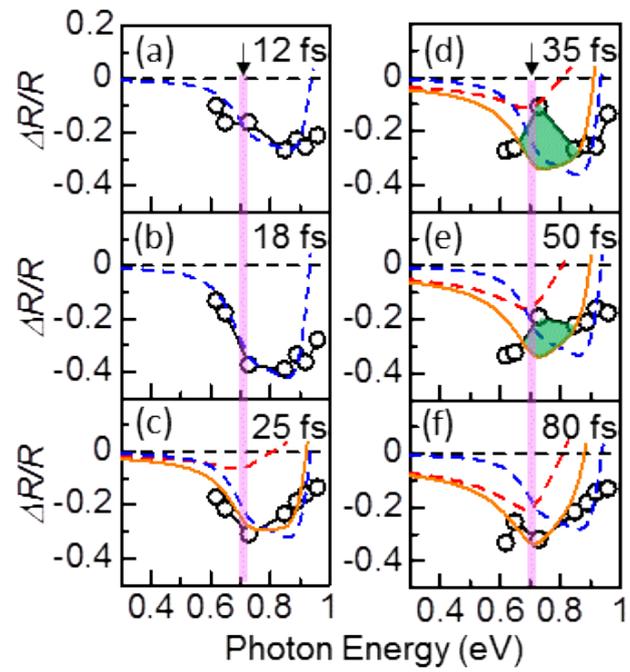

Kawakami et al. Fig. 17

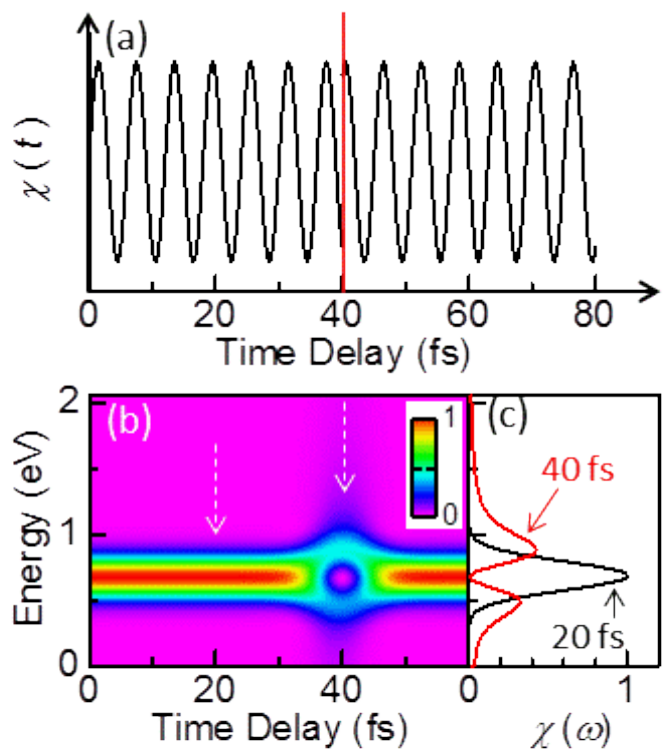

Kawakami et al.
Fig. 18

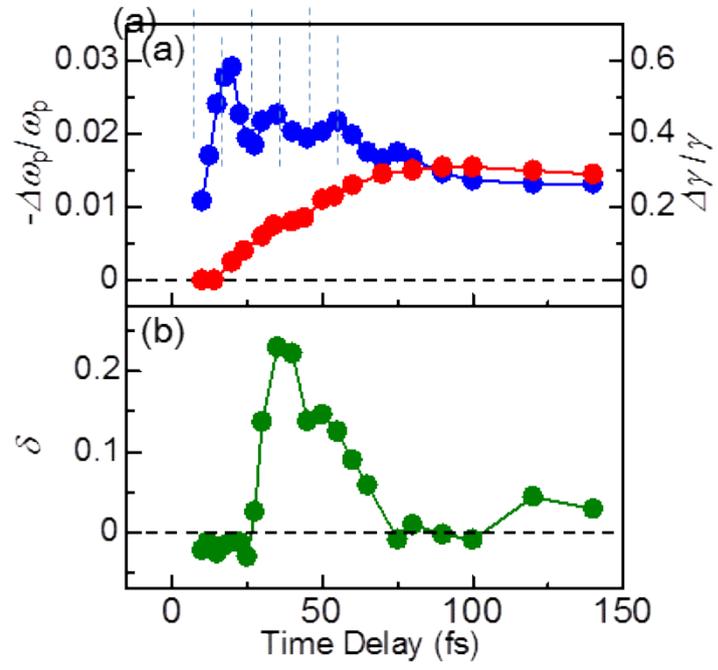

Kawakami et al.
Fig. 19